\documentclass[a4paper,15pt]{mn2e} 
\pdfoutput=1
\usepackage{graphicx}
\usepackage{deluxetable} 
\usepackage{lscape} 
\usepackage{epsfig} 
\usepackage{amssymb,amsmath}
\usepackage{cite}
\usepackage{csquotes}

\newcommand{\HI}{\mbox{H\,{\sc i}}} \newcommand{\MgII}{\mbox{Mg\,{\sc ii}}}
 \newcommand{\FeII}{\mbox{Fe\,{\sc ii}}}

\newcommand{\OI}{\mbox{O\,{\sc i}}}

\newcommand{\CaII}{\mbox{Ca\,{\sc ii}}}

\title[\CaII~Absorbers in the
SDSS]{\CaII~Absorbers in the Sloan Digital Sky Survey: Statistics} 
\author[G. M.Sardane et al.] {\parbox[t]{\textwidth}{\raggedright Gendith M. Sardane$^{1}$\thanks{E-mail:
gms48@pitt.edu}, David A. Turnshek$^{1}$, and Sandhya M. Rao$^{1}$} \vspace*{6pt}\\
$^{1}$Department of Physics and Astronomy and PITTsburgh Particle physics, Astrophysics, and
Cosmology Center (PITT PACC),\\ University of Pittsburgh, Pittsburgh, PA 15260\\ }

\begin{document}

\date{}

\pagerange{\pageref{firstpage}--\pageref{lastpage}} \pubyear{2014}
\maketitle

\label{firstpage}

\begin{abstract} We present the results of a survey for \CaII~$\lambda\lambda 3934,3969$
absorption-line systems culled from $\sim 95,000$ Sloan Digital Sky Survey (SDSS) Data Release 7 and 
Data Release 9 quasar spectra. With $435$ doublets identified in the catalog, this list is the largest
\CaII~catalog compiled to date, spanning redshifts $z < 1.34$, which corresponds to the most
recent $\sim 8.9$ Gyrs of the history of the Universe. We derive statistics on the \CaII~rest
equivalent width distribution and incidence (number density per unit redshift). We find that the
$\lambda 3934$ rest equivalent width ($W_{0}^{\lambda 3934}$) distribution cannot be described
by a single exponential function. A double exponential function is required to produce a
satisfactory description. The function can be written as a sum of weak and strong components: $
{\partial n}/{\partial W_{0}^{\lambda 3934}}= ({N_{wk}^{\star}}/{W_{wk}^{\star}}) exp({ -{W_{0}^{\lambda
3934}}/{W_{wk}^{\star} } }) + ({N_{str}^{\star}}/{W_{str}^{\star}}) exp({ -{W_{0}^{\lambda
3934}}/{W_{str}^{\star} } }) $. A maximum likelihood fit to the unbinned data indicates:
$N_{wk}^{\star}=0.140\pm0.029$, $W_{wk}^{\star}=0.165\pm 0.020~\textrm{\AA}$,
$N_{str}^{\star}=0.024\pm0.020$, and $W_{str}^{\star}=0.427\pm 0.101~\textrm{\AA}$. This suggests that
the \CaII~absorbers are composed of at least two distinct populations.  The incidence (product
of integrated absorber cross section and their co-moving number density) of the overall
\CaII~absorber population does not show evidence for evolution in the standard cosmology. 
The normalization of the no-evolution curve, i.e., the value of the \CaII\ incidence extrapolated to
 redshift $z=0$, for $W_{0}^{\lambda 3934} \ge 0.3$ \AA,  is $n_0=0.017 \pm 0.001$. 
In comparison to \MgII~surveys, we found that only $3\%$ of \MgII~systems in the SDSS have
\CaII, confirming that it is rare to identify \CaII~in quasar absorption-line surveys.
 We also report on some preliminary investigations of the nature of the two populations of \CaII~absorbers,
 and show that they can likely be distinguished using their \MgII~properties.

\end{abstract}


\section{Introduction} \label{intro} 

A successful and complete theory of galaxy formation and
evolution must not only explain the properties of the luminous components of galaxies, but also
account for the properties, kinematics, and evolution of gaseous structures associated with them. 
Quasar absorption lines (QALs) are an extremely powerful
 probe of the physical properties and kinematics of the gas in galactic, intergalactic and
circumgalactic environments. Since the detection of gaseous structures in absorption is
independent of the luminosity of the absorbing medium, quasar spectroscopy has been crucial in
providing a wealth of information on the distribution and evolution of matter in the Universe.
Without the selection bias caused by galaxy brightness and surface brightness limitations, one
can identify structures that are fainter than what traditional imaging studies allow.
QAL studies have resulted in the identification of a gamut of intervening gaseous absorbers from the
 coolest molecular clouds detected in H$_2$ (e.g., Noterdaeme et al. 2008) to the predominantly neutral 
 regions identified as \HI\ damped Lyman alpha systems (DLAs) and
low-ionization \MgII\ absorbers (e.g., Noterdaeme et al. 2012, Rao, Turnshek, \& Nestor 2006, Quider et al. 2011, Seyffert et al. 2013),
as well as hot ionized plasma in the extended halos of galaxies (e.g., Werk et al. 2014). 
The resonance transitions for the most common atoms and ions fall in the rest-frame ultraviolet (UV). 
Consequently, the QALs used to explore and make identifications of these various gaseous components
studies have often concentrated on absorbers at moderate to high redshifts where these lines 
fall at wavelengths accessible to optical ground-based telescopes. Given available time allocations,
the option of using space-based telescopes such as the Hubble Space Telescope (HST)
to perform large UV QAL surveys is impractical, expensive, and unrealistic. Consequently, 
large statistical studies of absorption line systems and the gaseous components of low-redshift galaxies that they trace are lacking. 

One particular class of absorber, which falls at optical wavelengths at low redshift, is that 
traced by the \CaII~H \& K doublet, i.e. \CaII~$\lambda\lambda3934,3969$.  It is a resonance doublet 
transition of singly ionized calcium from the ground state with rest-frame wavelengths $\lambda=3934.78~\mathrm{\AA}$ (\CaII~K) and
$\lambda=3969.60~\mathrm{\AA}$ (\CaII~H). The energy required to photoionize the neutral Ca atom
is $6.11~\mathrm{eV}$. However, the energy required to photoionize Ca$^{+}$ is only
$11.87~\mathrm{eV}$, a value that is slightly less than the ionization potential of \HI. Thus,
Ca$^+$~may not be the dominant ionization state of calcium. Moreover, Ca is a highly 
refractory element, being among the most depleted in the interstellar medium  (Savage and 
Sembach 1996; Wild and Hewett 2005; Wild, Hewett \& Pettini 2006).
Thus, \CaII\ is a rare class of absorber, which nevertheless is an important diagnostic of key
 physical properties of the gas such as its density, degree of self-shielding, and dust content.

Strong \CaII~absorption may preferentially arise in environments where some fraction of the dust
grains has been destroyed, and the proportion of gaseous Ca has been enhanced by a large factor due to
supernova-driven shocks associated with recent star-formation (Routly \& Spitzer 1952). More
recent studies of a handful of \CaII~absorbers (Wild \& Hewett 2005; Wild, et al. 2006; 
Wild, Hewett \& Pettini 2007; Nestor et al. 2008; Zych et al. 2007; Zych et al. 2009)
indicate that strong \CaII~systems preferentially reside in dense, dusty, neutral, metal-rich,
molecular $H_{2}$-bearing environments –- the reservoirs for subsequent star-formation.
Moreover, measurements (Zhu \& Menard 2013) of the average density profile of \CaII~gas around
galaxies out to $\mathrm{\sim200~kpc}$ using cross-correlation analysis of the positions of
$\sim 10^{6}$ foreground galaxies with $\sim 10^{5}$ background quasars in the Sloan
Digital Sky Survey (SDSS) concluded that most of the \CaII~in the Universe is in the circum-
and intergalactic environments, and that the \CaII~content in galaxy halos is larger for
galaxies with higher stellar mass and star formation rates.

Studies of the extent of rare \CaII~absorbers around galaxies will, therefore, place important
empirical contraints on models for the existence of cool gas in the extended regions of
galaxies. This includes models of cold accretion (e.g. Dekel \& Birnboim 2006, Kere$\check{\rm s}$ et al.
2009, Stewart et al. 2011, and references therein) and models relying on radiation pressure
driving from massive clusters followed by ram pressure driving from SNe (e.g, Nath \& Silk 2009,
Murray et al. 2011, Sharma \& Nath 2012, and references therein), which can launch cool gas out
beyond 50 kpc. These processes have implications for the fueling and evolution of galaxies
(Dav\'e, Oppenheimer \&  Finlator 2011; Dav\'e, Finlator \& Oppenheimer 2011, and references therein): cold accretion fuels star formation, while
resulting feedback and outflows quench it. Furthermore, such studies are useful in understanding
trends in the colors, luminosities, morphologies and orientations of galaxies, as well as the
dust-content and metal-enrichment of the IGM/CGM. Since \CaII\ can be detected in ground-based 
surveys down to $z=0$,  the lowest redshift \CaII\ systems allow 
for detailed studies of the absorbers and their host galaxy environments. 

In this paper, we present the results from the largest sample of \CaII~$\lambda\lambda 3934,
3969$ absorbers ever compiled. In a blind survey of
roughly $95,000$ quasar spectra from the Seventh and Ninth Data Release of the SDSS (SDSS-DR7, DR9), we
identified $435$ \CaII~doublets with $W_{0}^{\lambda 3934} \ge 0.15 ~\mathrm{\AA}$. The
wavelength coverage of the SDSS spectrum allows us to probe the redshift interval $z <
1.34$, which corresponds to $\sim8.9~\mathrm{Gyr}$ of cosmic history.
More importantly, \CaII~gives us ground-based access to the low redshift regime of
$z < 0.34$, equivalent to $\sim 4$ Gyrs of cosmic history, unlike any other commonly observed ionic
transition.

The paper is organized as follows: In \S2 we describe the data reduction process: the continuum
fitting and line-finding algorithms, the selection criteria we imposed, and tests for
systematic biases. We then present our main results in \S3, where we derive the
absorber rest equivalent width (REW) parametrization and evolution and the absorber redshift 
number density and its evolution, along with results on \CaII~doublet ratios and how the incidence of \CaII~absorbers 
compares with that of \MgII~absorbers. In \S4, we discuss evidence for two populations of \CaII~
absorbers. We summarize and present our conclusions in \S5.

Throughout the paper, we assume a standard $\mathrm{\Lambda}$CDM cosmology with
$\mathrm{H_{0}=71~km~s^{-1}Mpc^{-1}}$, $\mathrm{\Omega_{M}=0.27}$, and
$\mathrm{\Omega_{\Lambda}=0.73}$ (Spergel et al. 2007; Komatsu et al. 2011).

\section{The SDSS \CaII~Catalog} \label{obs} 

From its early beginnings, the SDSS (York et al. 2000) 
has been pivotal in advancing moderate resolution quasar absorption line spectroscopy by 
providing a huge increase in the number of quasar spectra available for absorption line surveys. 
Spectroscopy from the SDSS-I/II data releases has resulted in over 100,000 quasar spectra in the seventh data release (Schneider et al. 2010),
The spectra were obtained using a pair of similar multi-object fiber spectrographs mounted on a dedicated 2.5-m wide-field telescope.
Each spectrograph has $640$ three-arcsecond-diameter fibers, with a combined spectral coverage of $3800-9200~\mathrm{\AA}$.
More recently, the ninth data release provided an additional \text{$\sim 80,000$} quasar spectra
(Ahn et al. 2012; P\^aris et al. 2012) from $\sim 1.5$ years of data from the SDSS-III Baryon Oscillation 
Spectroscopic Survey (BOSS, Schlegel et al. 2007; Dawson et al. 2013). The improved BOSS spectrograph (Smee et
al. 2013) has $1000$ two-arcsecond-diameter fibers, and has an extended wavelength coverage 
of $3600-10,400~\mathrm{\AA}$. Both the SDSS and BOSS spectrographs have 
approximately the same resolution ranging from $1500$ at $3800~\mathrm{\AA}$ to
$2500$ at $9000~\mathrm{\AA}$. 

In this work, we utilize the most recent entries found in the SDSS DR7 and DR9 quasar 
catalogs of Schneider et al. (2010) and P\^aris et al. (2012), respectively.
The SDSS spectral coverage corresponds to an absorption redshift interval
of $z < 1.34$ in the \CaII~$\lambda3934$ absorption line. We confined our search for \CaII\ 
absorption lines in the BOSS data set to redshifts $z<1.34$ even though the BOSS quasar spectra have redshift 
coverage up to $z=1.64$. 

\subsection{Quasar Sample Selection}

\begin{figure*} 
\includegraphics[width=0.75\textwidth]{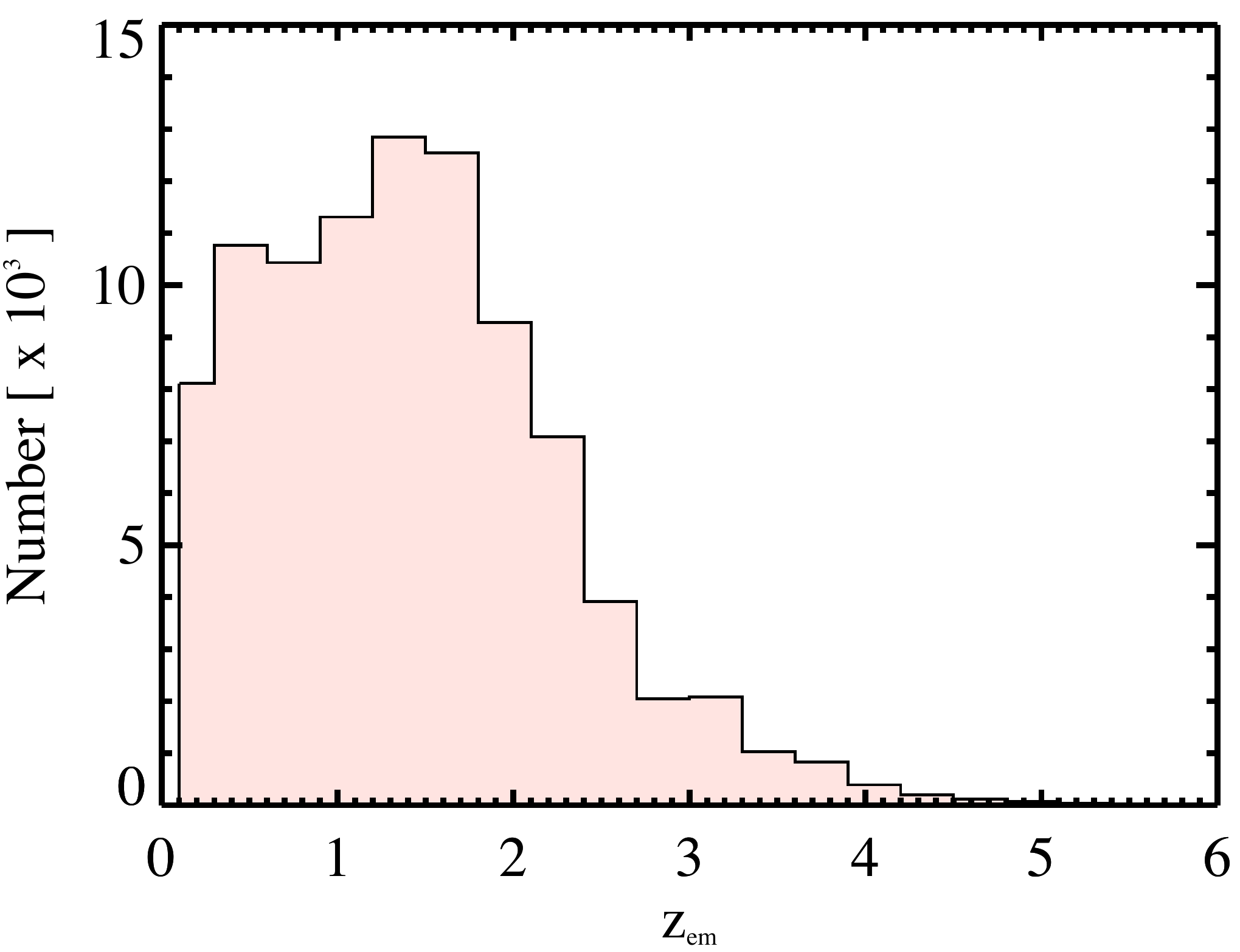} 
\caption{The distribution of emission redshifts of the SDSS quasars, with magnitudes 
$i < 20$ and $z_{em}\ge 0.1$, used to search for the \CaII~ $\mathrm{\lambda\lambda 3934,3969}$ absorption doublet. The
distribution has a mean of $<\!z_{em}\!> = 1.4$, and a maximum redshift of $z_{em} = 6.0$.}
\label{QSOzem} 
\end{figure*}

In order to ensure adequate signal-to-noise ratios, we restricted the quasar sample to
 SDSS fiber magnitudes of $i < 20$, and to minimize the incidence of
galaxies that have been misidentified as quasars, we only considered quasars with $z_{em} \ge 0.1$.
In addition, we searched for \CaII\ at wavelengths greater than
6000 km s$^{-1}$ redward of the quasar Ly-$\alpha$ emission line. The quasar with the highest 
redshift in our sample has $z_{em} = 6.0$. We used the
catalogs compiled by Shen et al. (2011) for SDSS DR7, and extended by P\^aris et al. (2012)
for SDSS DR9, to exclude broad absorption line quasars from our search. Altogether, 
$94,114$ quasar spectra were useful for the search of the \CaII\ absorption doublet.
The distribution of emission redshifts, $z_{em}$, of the quasar sample 
is shown in Figure \ref{QSOzem}. The distribution has a mean of $<\!z_{em}\!> = 1.4$.

\subsection{Data Reduction} 

The construction of our \CaII~absorber sample closely follows the methods adopted for the construction of the 
University of Pittsburgh SDSS \MgII\ catalog described in Nestor et al. (2005), Rimoldini (2007), and Quider et al. (2011). Quite
generally, our data reduction proceeds three-fold as follows: (1) automated quasar
processing, (2) visual inspection of the automatically flagged doublet candidates, and (3) 
measurement of the line strengths of the doublets that passed the stringent visual inspection.

\begin{figure*} 
\includegraphics[width=0.75\textwidth]{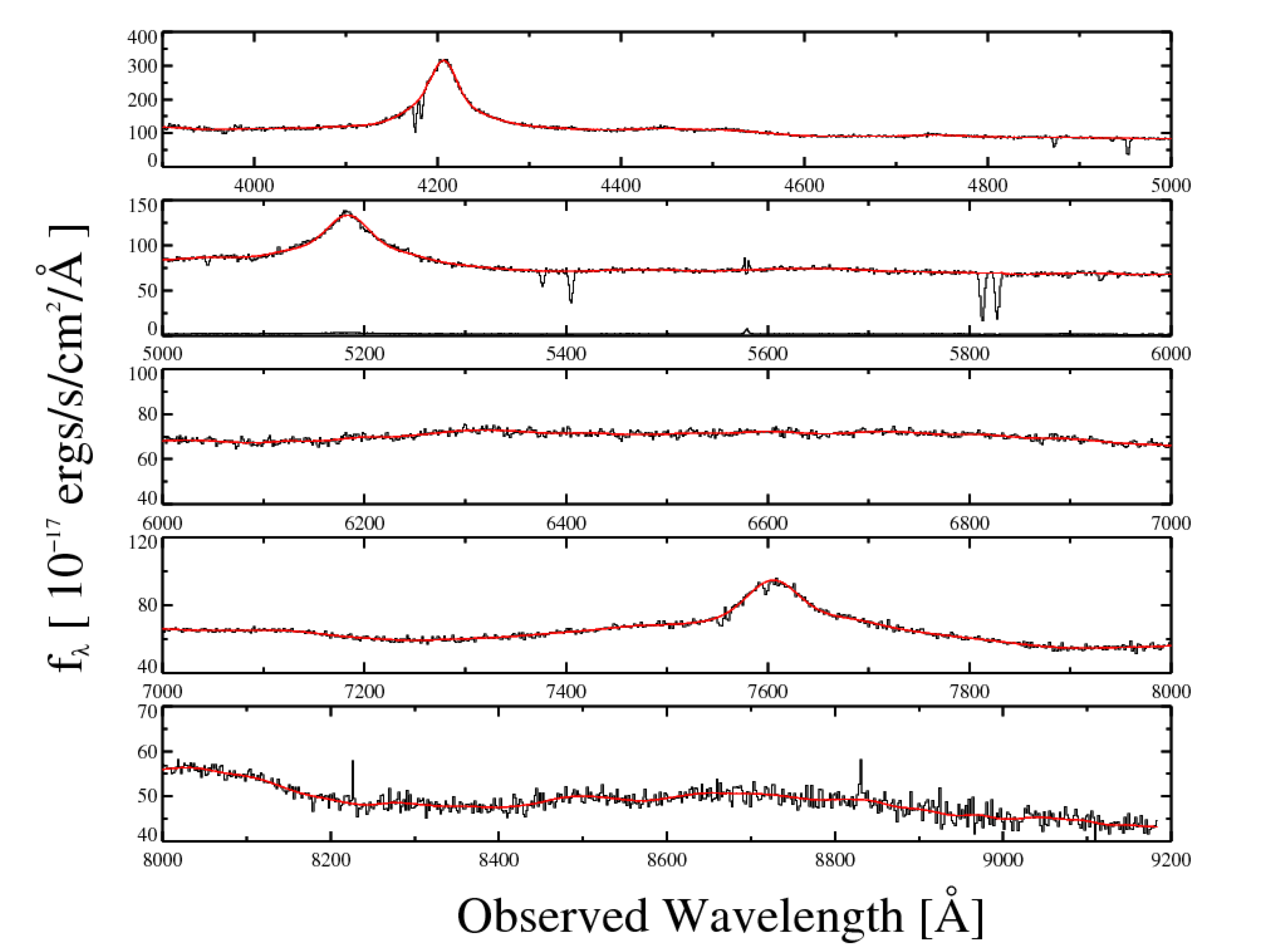} 
\caption{An example quasar
spectrum with the psuedo-continuum fit overplotted in red. In this example, \text{$z_{em} =
1.720$}, and the median error is $\sim2.6\%$~of the flux. } 
\label{continuum}
\end{figure*}

\begin{figure*} 
\includegraphics[width=0.75\textwidth]{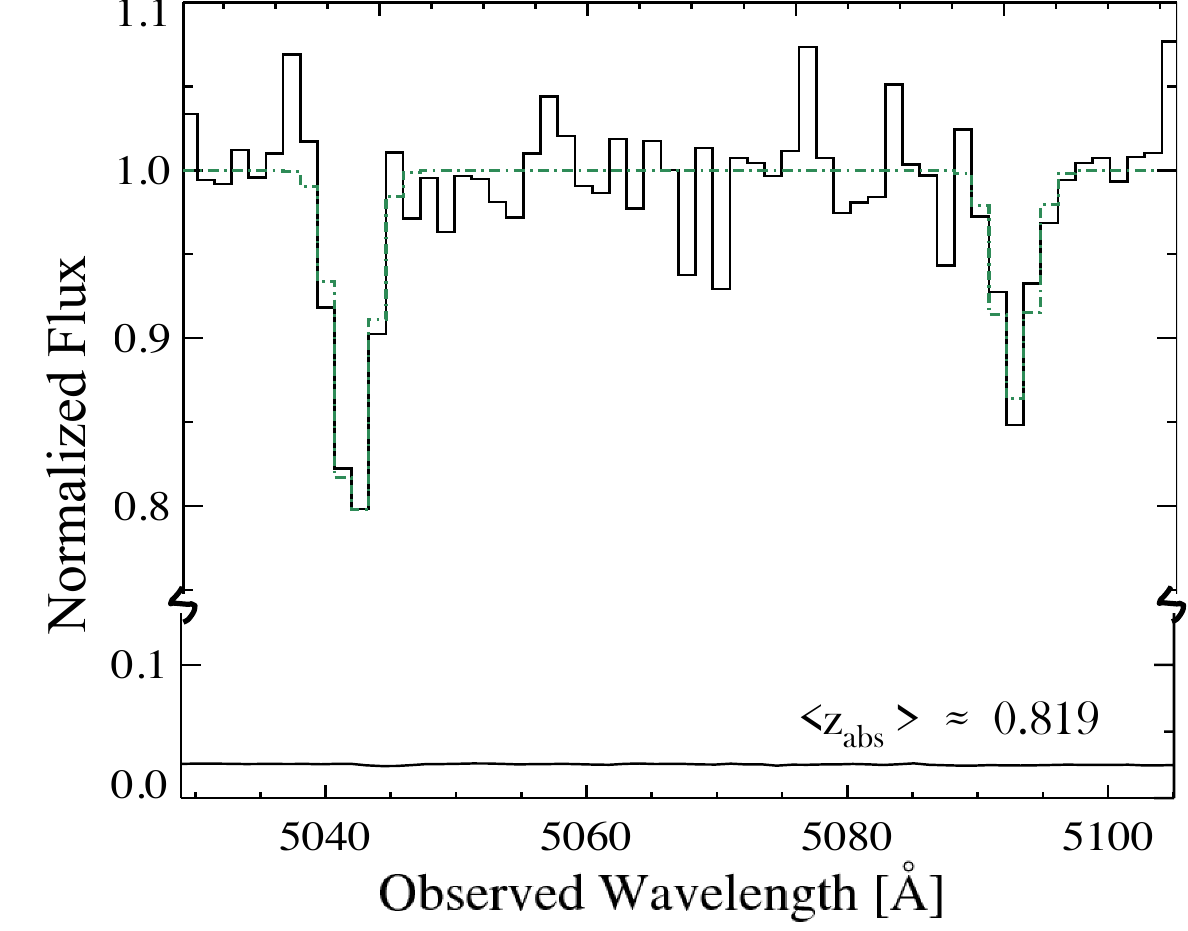} 
\caption{An example \CaII~absorber system at $z_{abs} = 0.819$. The green dot-dash curve is a
double Gaussian profile that was fit simultaneously to both members of the doublet. Note that to
emphasize the feature, as well as its error array, the spectrum is truncated between $\sim 0.1-0.8$
in normalized flux units. }
\label{exampleAbs} 
\end{figure*}

The automated processing procedure consisted of two stages, i.e.,  the pseudo-continuum
normalization and the search for \CaII~doublets. A combination of cubic splines and
Gaussians were employed to determine the pseudo-continuum fit, which included both the true
continuum as well as the broad emission features. In Figure \ref{continuum}, we show an example spectrum with
the pseudo-continuum fit overplotted in red. For the vast majority of spectra, the
continuum-fitter worked quite well, even in regions of poor signal-to-noise ratios. The error in the
normalized flux is derived by dividing the flux error array by the fitted continuum. We do not
propagate any errors in the continuum fit to determine the normalized error array. However, we will later show that a
$20\%$ error in continuum level determination is not the dominant source of uncertainty; given the 
relatively small number of detected systems, statistical Poisson errors are still the major source of uncertainty.
  
All normalized spectra are then searched for \CaII\ absorption using a line-finding algorithm that
flags possible \CaII\ candidate doublets based on the doublet separation and line significance
levels. To isolate a sample of intervening absorbers, we only accepted candidates that were
separated in velocity by at least~ $6000~\mathrm{km~s^{-1}}$ from the quasar emission
redshift, and from $z=0$. Thus, biases that could arise due to an over-density of absorbers in
the vicinity of quasar environments and the Milky Way, respectively, were minimized.

All candidate doublets were then visually inspected to check for satisfactory continuum fits,
blends, and potential false detections due to the presence of absorption lines at other redshifts that 
could mimic the \CaII\ profile. We further note that we painstakingly examined each absorption
feature flagged by the line-finding routine and retained systems after eliminating every other possibility.
In order to identify a \CaII\ system, we required the detection of the $\lambda3934$ line {\it and}
 the $\lambda3969$ doublet partner. We
required a $5\sigma$ minimum level of significance for  the $\lambda3934$ line
and a $2.5\sigma$ minimum level of significance 
for the $\lambda 3969$ doublet partner. From their oscillator strengths, $f=0.682$ for the 
$\lambda3934$ line and $f=0.330$ for the $\lambda3969$ line (Kramida et al. 2013),
the secondary $\lambda3969$ line is expected to be roughly half as weak as the primary
$\lambda3934$ line in the unsaturated regime. We measured the doublet REWs, $W_0^{\lambda3934}$ and $W_0^{\lambda3969}$, by
fitting unresolved Gaussian profiles to both lines simultaneously, with full width half maxima (FWHM)
 given by the resolution of the SDSS spectrograph. Candidates with doublet ratios (DRs) that were outside the
physically allowed range of $1.0-\sigma_{DR} \le W_0^{\lambda3934}/W_0^{\lambda3969} \le 2.0 +
\sigma_{DR}$ were eliminated. The error in the doublet ratio, $\sigma_{DR}$, was estimated
assuming Gaussian uncertainties. The redshift of an absorber was determined from the weighted average
of the wavelength centroids of the two fitted doublet Gaussian profiles. 
Figure \ref{exampleAbs} shows an example absorber that passed our selection cuts.

\begin{figure*} \includegraphics[width=0.75\textwidth]{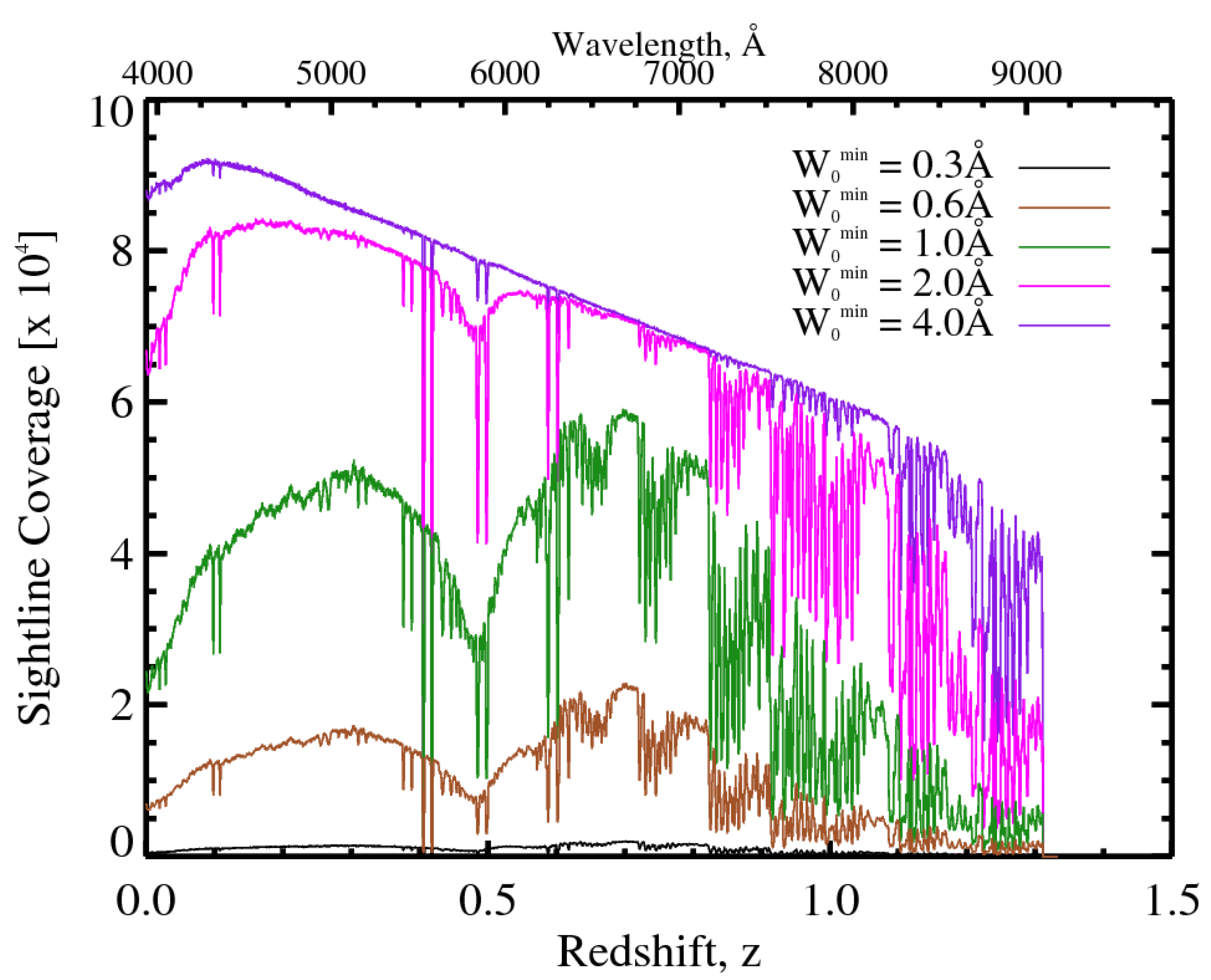} 
\caption{Sightline coverage 
 for the \CaII~survey in the SDSS DR9 as a function of absorber
redshift and REW threshold: \text{$\mathrm{W_0^{min} = 0.3, 0.6, 1.0, 2.0,
4.0~\mathrm{\AA}}$}. This gives the total number of lines of sight with sufficient
signal-to-noise ratio to detect at least a saturated \CaII~doublet at the \text{$5\sigma
$},\text{$2.5\sigma$} significance levels. See text. The wide, deep feature near $z=0.5$ is due to the 
SDSS dichroic. The sharp narrow dip in the middle of the dichroic is from the prominent
\OI~$\lambda 5578$~night sky line. The series of sharp declines 
redward of $\mathrm{z \approx 0.8}$ are the due to strong night sky lines. The 
additional constraint on the
$\lambda{3969}$~line in Eq. \ref{sensitivityEq} results in the doubling of narrow
dips.} \label{coverage} \end{figure*}

The survey sightline coverage, or sensitivity function, is shown in Figure \ref{coverage} as a function of absorber
redshift and minimum detectable REW threshold, $W_0^{min}$. The corresponding SDSS wavelength coverage is indicated 
by the top axis. The sightline coverage is the total number of lines of sight with sufficient signal-to-noise ratio 
to detect the $\lambda 3934$ \AA\ line with $W_0^{\lambda 3934} \ge W_{0}^{min}$ at a $\ge5\sigma$ level of significance, 
\textit{and, at the same time,} detect the $\lambda 3969$ \AA\ line at a $\ge2.5\sigma$ level of significance.
We emphasize that in order to be included in the accounting of the total survey path, a single redshift 
pixel and its corresponding doublet pixel position must have sufficient signal-to-noise ratios to detect the doublet 
pair at significance levels of 5$\sigma$ and 2.5$\sigma$, respectively, and to detect both lines within the 
physically allowable doublet ratio range of $1.0-\sigma_{DR} \le W_0^{\lambda3934}/W_0^{\lambda3969} \le 2.0 + \sigma_{DR}$.
\footnote{We note that no previous survey for absorption line doublets has imposed as stringent a doublet-finding 
criterion as employed here. Past surveys imposed a significance cut only on the stronger member of the doublet.} 
The strongest possible $\lambda3969$ absorption line is given by a profile with $DR=1.0$, therefore,
a pixel is rejected from the redshift path if a saturated
$W_0^{\lambda 3969}$ line, where $W_{0,min}^{\lambda3934}=W_{0,min}^{\lambda3969}$,  cannot be detected
at this position with a significance level of at least $2.5\sigma$. The sensitivity function for our
survey is given by Equation \ref{sensitivityEq}:

\begin{equation} 
\begin{split} g(W_{0}^{\lambda 3934},z) = \sum_{i=1}^{N_{LOS}}H(z-z_{min(i)})
H(z_{max(i)}-z) \\ \times H[W_{0}^{\lambda 3934} - 5\sigma_{0}(z) ]H[W_{0}^{\lambda 3969} -
2.5\sigma_{0}(z) ] \end{split} \label{sensitivityEq} \end{equation}

\noindent where the sum is over the total number of lines of sight, $N_{LOS}$, and $H$ is the Heaviside function.
Using $\lambda_{min}$ and  $\lambda_{max}$ to
indicate the wavelength limit of each quasar spectrum, we write the minimum (maximum)
redshift coverage, $z_{min}$ ($z_{max}$), for each quasar spectrum as

\begin{equation} 
z_{min} = \begin{cases} 0.02 & \text{if $\lambda_{0} \ge \lambda_{min}$}\\
\lambda_{min} / \lambda_{0} - 1 & \text{if $\lambda_{0} < \lambda_{min}$} \end{cases}
\end{equation}

\begin{equation} 
z_{max} = \begin{cases} z_{em} - 0.02 & \text{if $\lambda_{0}(1+z_{em}) <
\lambda_{max}$}\\ \lambda_{max} / \lambda_{0} - 1 & \text{if $\lambda_{0}(1+z_{em}) \ge
\lambda_{max}$}. \end{cases} 
\end{equation}

The prominent deep feature occurring near absorber redshift $z \sim 0.5$ (\text{$6000
~\mathrm{\AA}$}) is due to the dichroic (Schneider et al. 2010). The conspicuous
 absorption features redward of $z \sim 0.8$ are due to strong night sky lines in many spectra.
 Taking into account the line significance of both members of
the \CaII~doublet in calculating redshift path results in the doubling of
 narrow dips seen in Figure \ref{coverage}, as one might expect.

\begin{figure*} 
\includegraphics[width=0.75\textwidth]{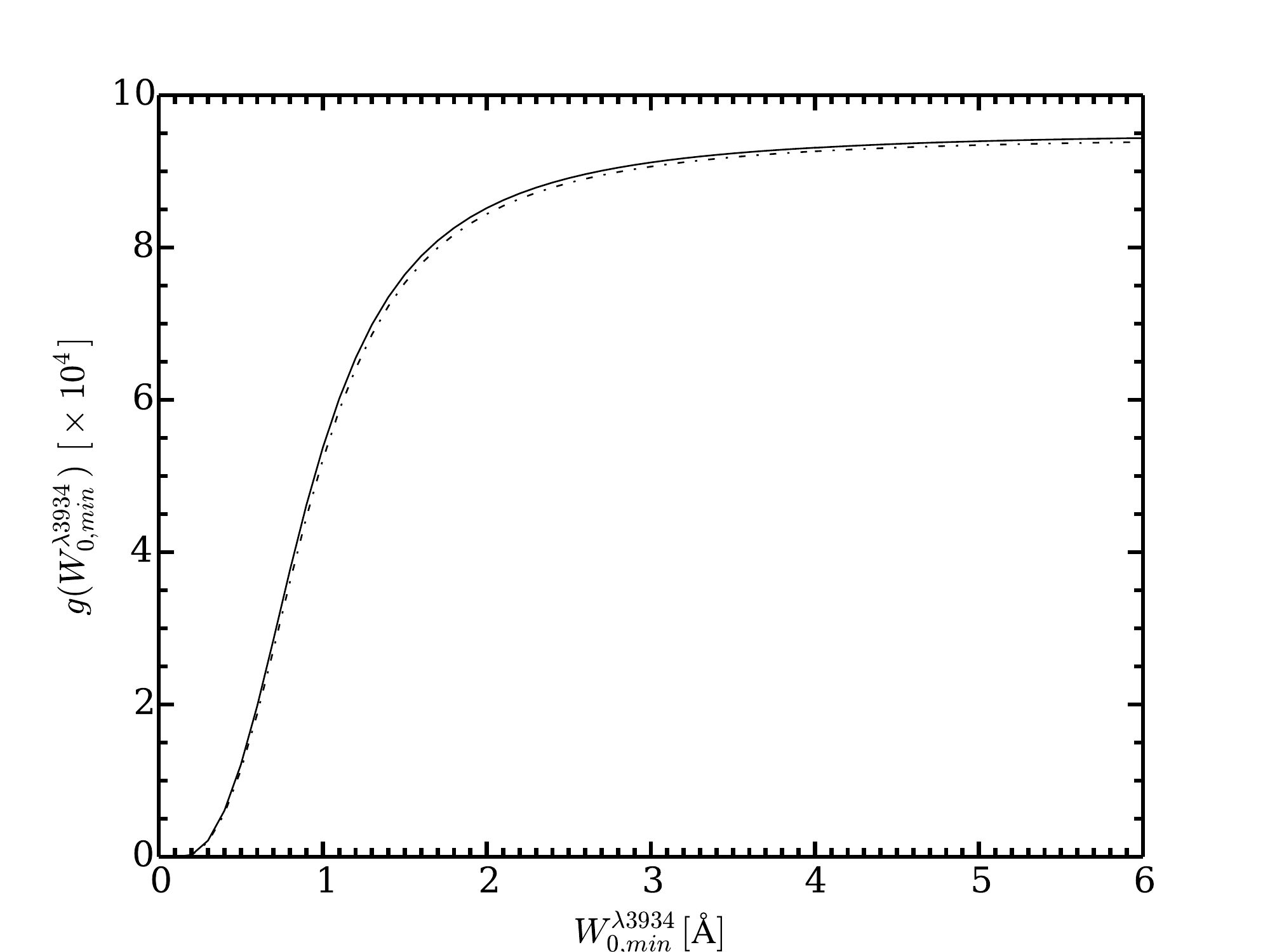} 
\caption{The cumulative
pathlength, $g(W)$, as a function of REW threshold is shown here as the black
solid curve. The decrease in the path due to an additional $20\%$ uncertainty in the
pseudo-continuum fit added in quadrature is shown by the black dash-dot curve. The difference
between the two cumulative path lengths peaks at
$8\%$ at $W_{0, min}^{\lambda 3934} = 0.2~\mathrm{\AA}$, decreases to $2\%$ at
$W_{0,min}^{\lambda 3934} = 1.0~\mathrm{\AA}$, and to 0.04\% 
at $W_{min,0}^{\lambda3934}=6.0~\mathrm{\AA}$.} 
\label{cumulative} 
\end{figure*}

Integrating the sensitivity function of Figure \ref{coverage} over the allowed redshift interval
for \CaII, as determined by each SDSS spectrum, gives the cumulative path length of the survey,
$g(W_0^{min})$, as a function of REW threshold $W_0^{min}$. The solid black curve shown in
Figure \ref{cumulative} effectively describes the sensitivity of
the survey to the measurement of a given strength of the $\lambda3934$ line. For example, 
 a $\lambda3934$ line with REW $W_0^{\lambda3934}\ge 1.0$~\AA\ can only be detected in roughly half of
the total available sightlines. The sensitivity then asymptotes to $94,114$ lines of sight
at large REWs. The dash-dot black curve depicts the decrease in the cumulative path that
would result from a (conservativley chosen) $20\%$ error in the continuum fit, added to the flux error array in
quadrature. The difference between the two paths peaks at $8\%$
at $W_{min,0}^{\lambda3934}=0.2~\mathrm{\AA}$, decreases to $2\%$ at
$W_{min,0}^{\lambda3934}=1.0~\mathrm{\AA}$, and to 0.04\% 
at $W_{min,0}^{\lambda3934}=6.0~\mathrm{\AA}$.

\subsection{Monte Carlo Simulations to Determine False Positives and Systematics}
We ran Monte Carlo simulations of the absorber catalog to test the efficiency of our detection
routine and identify possible biases and systematic effects. Prior to the simulations, we masked
out all detected \CaII~systems from their respective spectra, and used the edited spectra for the simulations
instead. Using the observed distributions for the absorption redshift, $W_0^{\lambda 3934}$, and FWHM of the absorbers, and
a uniformly distributed doublet ratio, we generated \text{$10,000$} \CaII~doublets and inserted
them into randomly-selected spectra. Approximately $7300$ of these appeared in regions of spectra with
sufficient signal-to-noise ratio that met our criteria for detection. We then ran the entire
data pipeline and recovered $97.7\%$ of these simulated doublets. Thus, we may have missed a 
maximum of 10 doublets in our search. In addition, no \CaII\ doublet 
that was not in the input list was falsely included in the output list of the simulation. 

\section{Results}

\subsection{The $W_{0}^{\lambda 3934}$ Distribution}

\begin{figure*} 
\includegraphics[width=0.75\textwidth]{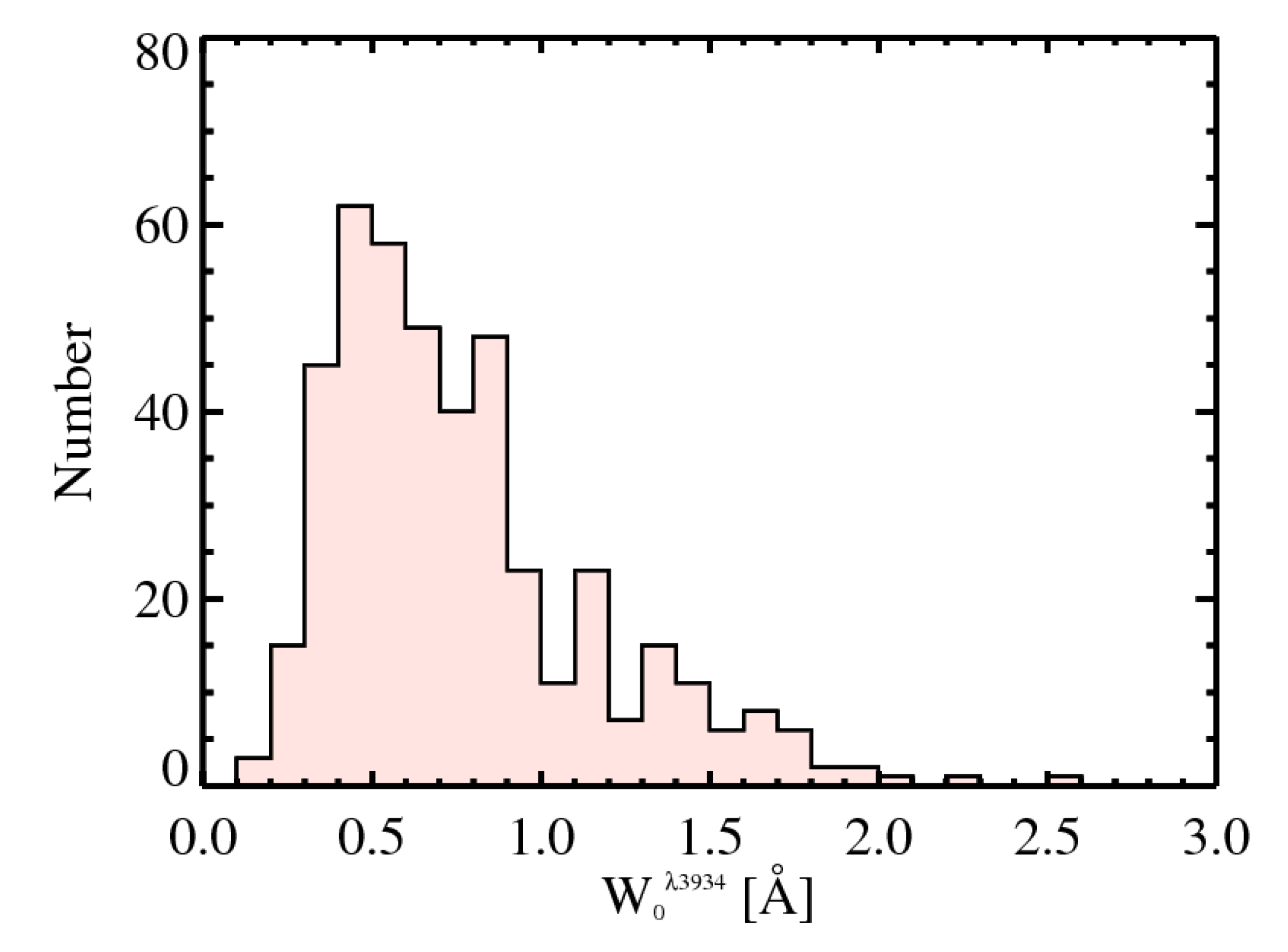} 
\caption{The observed REW distribution for $W_0^{\lambda 3934}$. The distribution has a mean of
$\mathrm{0.769 \AA}$ and a spread of 0.393 \AA. Measured REWs range from 0.163 \AA\ $\le W_{0}^{\lambda 3934}
\le 2.573$ \AA. } 
\label{rewdist} \end{figure*}

\begin{figure*} 
\includegraphics[width=0.75\textwidth]{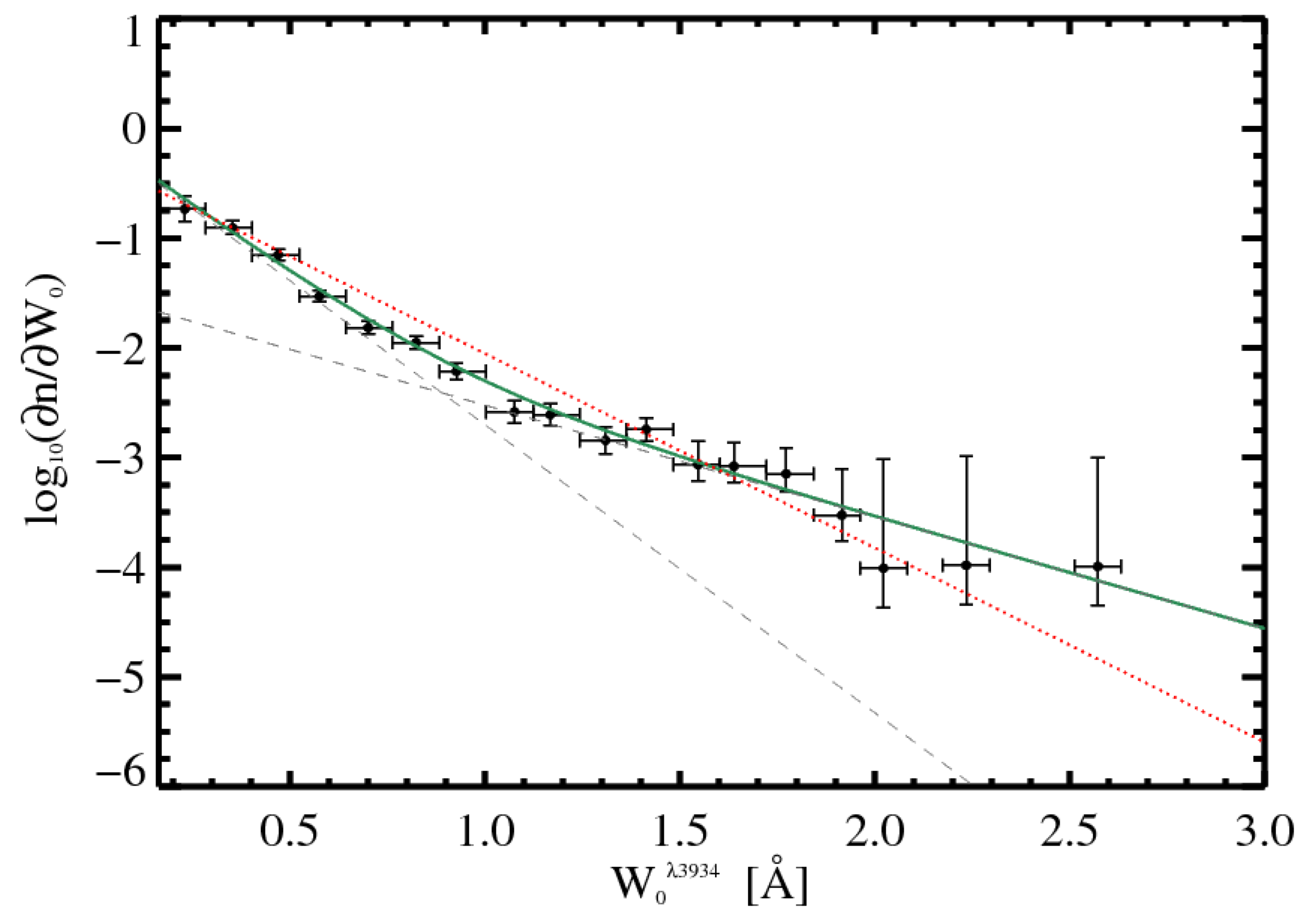} 
\caption{The sensitivity-corrected $W_0^{\lambda 3934}$ distribution. The double exponential model, Equation \ref{modelDist}, that maximizes the likelihood to
the unbinned data, is shown in green. The two single-exponential components of Equation
\ref{modelDist}~are plotted as grey dashed lines. The maximum likelihood fit using a
single exponential model is shown as the red dotted line. } 
\label{truerewdist} \end{figure*}

\begin{table*}
\caption{The \CaII\ Sample\tablenotemark{a}}
\begin{tabular}{cccccccc}
\hline
\hline

Quasar 	&	SDSS 	& $z_{em}$ & $z_{abs}$ & $W_0^{\lambda3934}$ & $\sigma(W_0^{\lambda3934})$	& $W_0^{\lambda3969}$ &	$\sigma(W_0^{\lambda3969})$\\
	& g mag & & &($\mathrm{\AA}$) & ($\mathrm{\AA}$) & ($\mathrm{\AA}$)	& ($\mathrm{\AA}$)\\
\hline

J001214.19$-$095922.9 & 19.44 & 1.262 & 0.6901 & 0.773 & 0.137 & 0.412 & 0.112\\
J001444.02$-$000018.5 & 17.95 & 1.550 & 0.0277 & 0.326 & 0.056 & 0.201 & 0.055\\
J002940.02+010528.5 & 17.83 & 1.388 & 0.3732 & 0.302 & 0.059 & 0.213 & 0.060\\
J004130.97+024222.5 & 18.81 & 2.308 & 0.7095 & 0.720 & 0.143 & 0.370 & 0.120\\
J004800.50+022514.9 & 18.96 & 2.160 & 0.5982 & 0.594 & 0.101 & 0.297 & 0.096\\
\hline
\noindent
\tablenotetext{a}{The table is available in its entirety online.}
\end{tabular}
\end{table*}

We identified $\mathrm{435}$ \CaII~doublets with $W_0^{\lambda 3934} >
0.160 ~\mathrm{\AA}$ and $z\lesssim1.34$. 
The first few entries of our \CaII\ catalog are presented in Table 1. The table is available in its entirety online. 
The observed $W_0^{\lambda 3934}$ distribution is shown in Figure
\ref{rewdist}. The strongest system we found has $W_0^{\lambda3934}=2.573~\mathrm{\AA}$, while
the weakest system has $W_0^{\lambda3934}=0.163~\mathrm{\AA}$. The distribution has a mean of
$<\!W_0^{\lambda3934}\!> = 0.769~\mathrm{\AA}$ and a spread of $\sigma=0.393~\mathrm{\AA}$. 
Combining the observed
distribution from Figure \ref{rewdist} with the sensitivity function in Figure \ref{cumulative},
we obtain the sensitivity-corrected distribution for $W_0^{\lambda 3934}$, shown as the
binned data points in Figure \ref{truerewdist}. The errors are determined using Poisson 
statistics. Similar to what has been found for other classes of QAL systems, the
REW distribution rises with decreasing REWs.  However, the data clearly show a change in 
the slope of the distribution near  $W_0^{\lambda 3934} \approx 0.9$ \AA. 
The best-fit single exponential function, determined using a maximum likelihood estimate (MLE) on
 the unbinned data,  is clearly a poor fit as shown by  the red dotted line in Figure \ref{truerewdist}.
Therefore, we used a double-exponential function, written as the sum of weak and strong components,
to obtain a satisfactory fit. Equation \ref{modelDist} parametrizes the model with two
characteristic REWs, a $W_{wk}^\star$ and $W_{str}^\star$, and two normalization constants,
$N_{wk}^\star$ and $N_{str}^\star$, for the weak and strong components, respectively.

\begin{equation} 
\frac{\partial n}{\partial W_{0}^{\lambda 3934}} = \frac
{N_{wk}^{\star}}{W_{wk}^{\star}} e^{ - \frac{W_{0}^{\lambda 3934}}{W_{wk}^{\star} } } + \frac
{N_{str}^{\star}}{W_{str}^{\star}} e^{ - \frac{W_{0}^{\lambda 3934}}{W_{str}^{\star} } } 
\label{modelDist}
\end{equation}

The resulting fit parameters are $N_{wk}^{\star}=0.140\pm0.029$ and
$W_{wk}^{\star}=0.165\pm 0.020 \textrm{\AA}$ for the weak component, and 
$N_{str}^{\star}=0.024\pm0.020$ and $W_{str}^{\star}=0.427\pm 0.101 \textrm{\AA}$ for the strong component. 
The solid green curve shows this best-fit double exponential function MLE fit to the unbinned data. 
The dashed grey lines are the two individual components; from this fit, we determined that the change in slope
occurs at $W_0^{\lambda 3934}=0.88\mathrm{\AA}$.
 
 We evaluated the Akaike Information Criterion (AIC) for both the single and double exponential fits, and
obtained $AIC=46$ and $AIC=11$ for the two fits, respectively. The AIC is a measure
of the quality of the candidate models relative to each other (Liddle 2007). It is based on information
entropy and quantifies the trade-off between goodness of fit and complexity (i.e., the
number of parameters) of the model. Given a set of candidate models, the model with the smallest
AIC value has the strongest support. Thus based on the AIC values above, the data are significantly
better represented by  the double exponential fit. The existence of a change in slope 
and  the success of the double exponential model may be interpreted as
evidence for the existence of more than one class of \CaII\ absorber. We will address this possibility 
further in \S4.

\subsubsection{Redshift Evolution of $\partial n/\partial W_0$}

\begin{figure*} 
\includegraphics[width=0.75\textwidth]{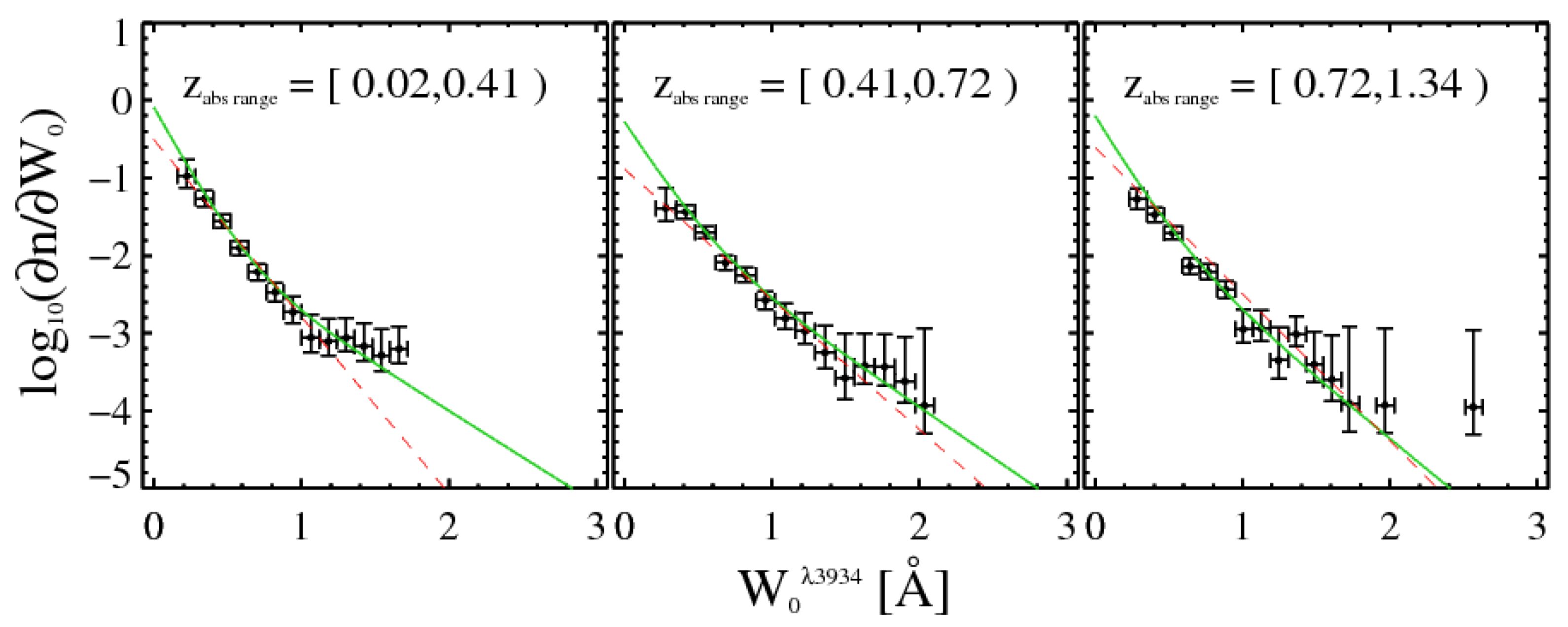} 
\caption{The
sensitivity-corrected $W_0^{\lambda 3934}$ distribution binned into three redshift intervals.
 The double exponential model (Equation
\ref{modelDist}) that maximizes the likelihood to the unbinned data is shown in green. The single
exponential fit is shown as the red dashed line. The single exponential model is
less favored over the double exponential model in all redshift intervals.} 
\label{binneddNdW} \end{figure*}

We now investigate the redshift evolution of the REW distribution, $\partial n/\partial W_0$. We
binned the data into three redshift subsamples, with each $z_{abs}$ interval having roughly
the same number of absorbers. The results are shown in Figure \ref{binneddNdW}. The solid green
curves show the MLE fit to the unbinned data. For comparison, the single exponential               
fits are shown as red dashed lines. The subsample in the lowest redshift bin
shows the most prominent departure from a single exponential fit. While the other two
subsamples still show some hints of a change in slope, it is less apparent given the increasing
size of the error bars at larger $W_0^{\lambda 3934}$. The AIC values in each subsample
suggest that the single exponential model is still less favored, although the degree of support has
decreased in the higher-redshift subsamples. Thus, we cannot discount
the possibility of the persistence of multiple populations across the different redshifts.

\begin{figure*} 
\includegraphics[width=0.75\textwidth]{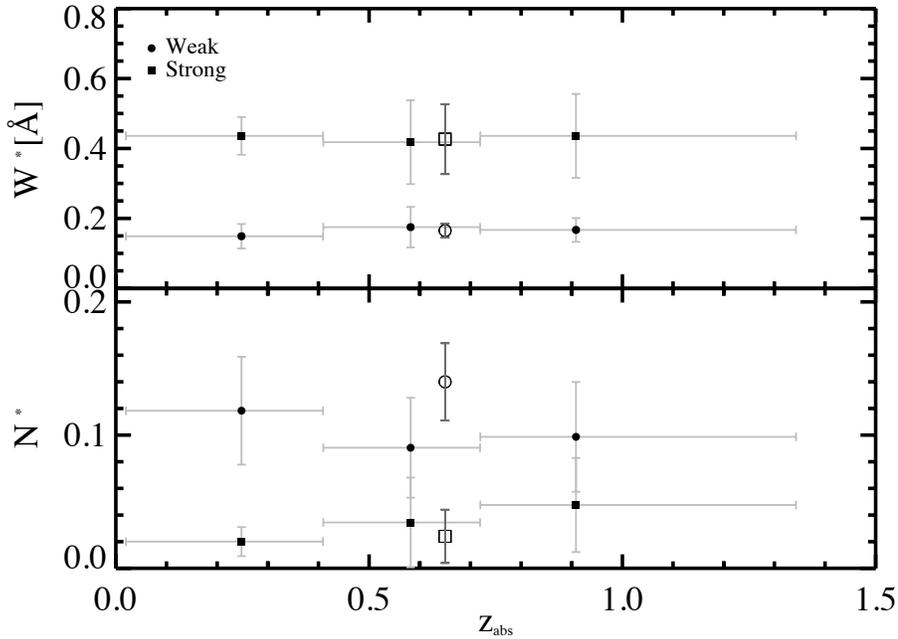} 
\caption{\textit{Top:}
The MLE characteristic REWs, $W^{\star}$, for the weak (filled circles) and strong (filled squares)
components, in the three
redshift ranges shown in Figure \ref{binneddNdW}. The error bars and bin sizes are shown in grey. The
$W^{\star}$ parameters imply a lack of redshift evolution in the slopes of each component.
\textit{Bottom:} The MLE normalizations for the two-component fit, which within the errors, are
also consistent with no evolution. For reference, we have plotted the results for the entire
sample as open circles and squares, which for clarity, are plotted at a location slightly offset from the
median $z_{abs}$.} 
\label{ParamsEvol} \end{figure*}

In Figure \ref{ParamsEvol} we plot the resulting double exponential model parameters $W_{wk}^\star$
and $W_{str}^\star$, and $N_{wk}^\star$ and $N_{str}^\star$, as a function of the mean 
$z_{abs}$ in each subsample. The result for the
entire \CaII~sample is shown as the open data points, plotted at a $z_{abs}$ that is 
slightly offset from the median for clarity. These plots clearly show that there is no evidence for
evolution in the shape of the exponenial distributions for either the weak or strong components of the fit. 
 In addition, Kolmogorov-Smirnov (KS) tests are also consistent with no evolution.

\subsection{The \CaII~Absorber Redshift Distribution}

\begin{figure*} \includegraphics[width=0.75\textwidth]{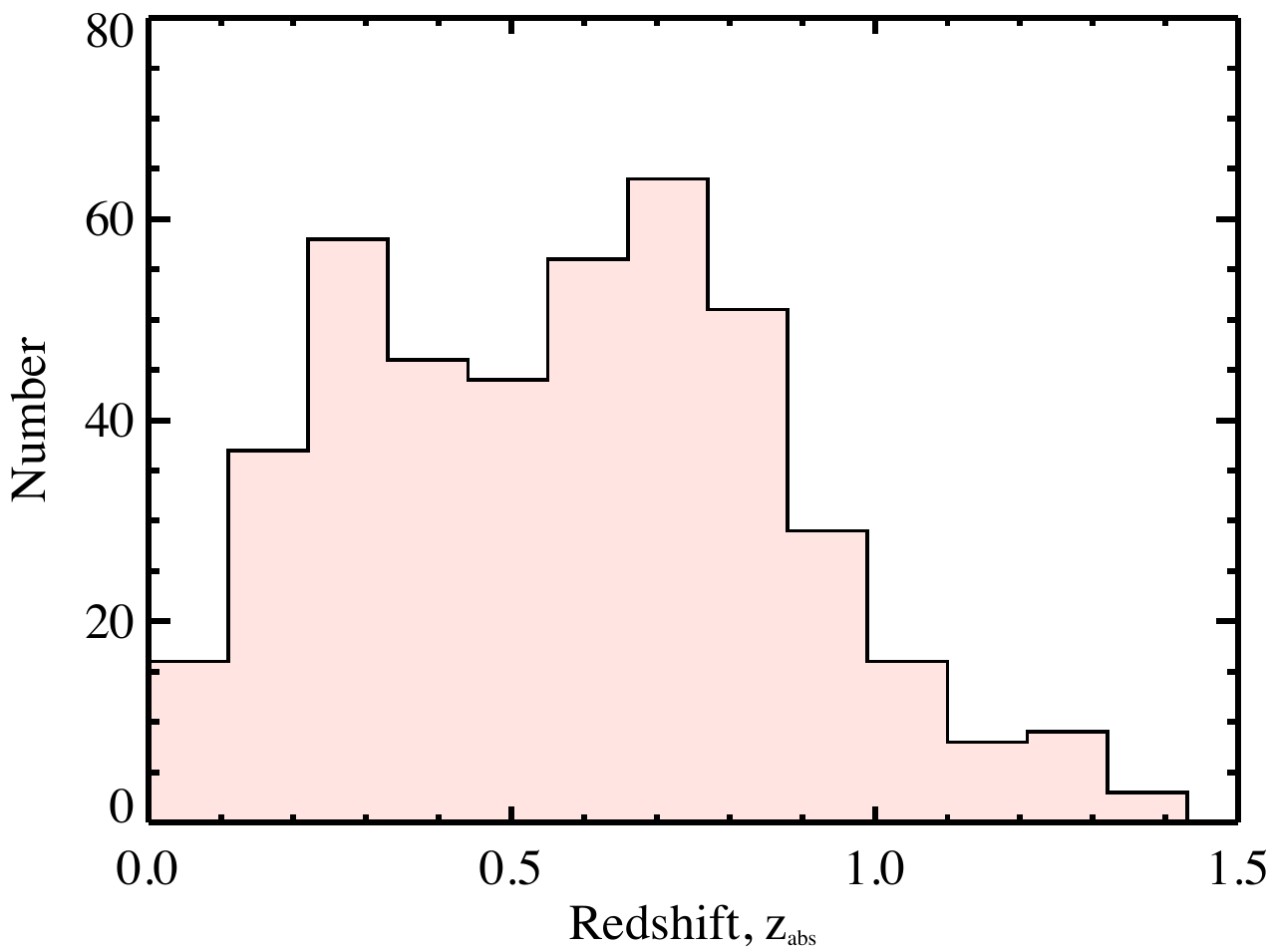} 
\caption{The observed
absorber redshift distribution shown in bins of $\Delta z = 0.1$, with mean $z_{abs}
= 0.579$ and standard deviation $\sigma = 0.296$. The poor SNR of SDSS spectra near
$z_{abs} \sim 0.5$, due to the SDSS dichroic, causes the decrease in the number of detected
\CaII~systems that pass our selection criteria at this redshift.} \label{zdist} 
\end{figure*}

The observed absorber redshift distribution is shown in Figure \ref{zdist}. 
As mentioned previously, SDSS spectra can be used to search for \CaII\ at redshifts 
$z_{abs} \lesssim 1.34$, equivalent to a lookback time of $ t_{LB} = 8.9 ~\mathrm{Gyrs}$, or
$\gtrsim 60\%$ of our cosmic history. 
The observed distribution has a mean redshift of $<\!z_{abs}\!> = 0.579$, and standard deviation of
$\sigma_{z_{abs}}=0.296$. The distinct drop in sensitivity near $z\sim0.5$ is mainly due to the
SDSS dichroic. 

The number density, $\partial n/ \partial z$, or the incidence of lines that have
$W_0^{\lambda3934}$ larger than a specified threshold $W_0^{min}$ over some redshift interval, is
given by

\begin{equation} 
\frac{\partial n}{\partial z}\Bigg \vert_{W_{0} > W_0^{min}} =
\sum_{\substack{W_{0,i} > W_{0,min}\\ z_{i} \in (z, z+dz) }} \frac{1}{g(W_{0,i}, z_{i})dz}
\end{equation}

\noindent whose variance is given by

\begin{equation} 
\sigma^2 = \sum_{\substack{W_{0,i} > W^0_{min}\\ z_{i} \in (z, z+dz) }}
\left(\frac{1}{g(W_{0,i}, z_{i})dz}\right)^{2}.
\end{equation} 

\begin{figure*} \includegraphics[width=0.75\textwidth]{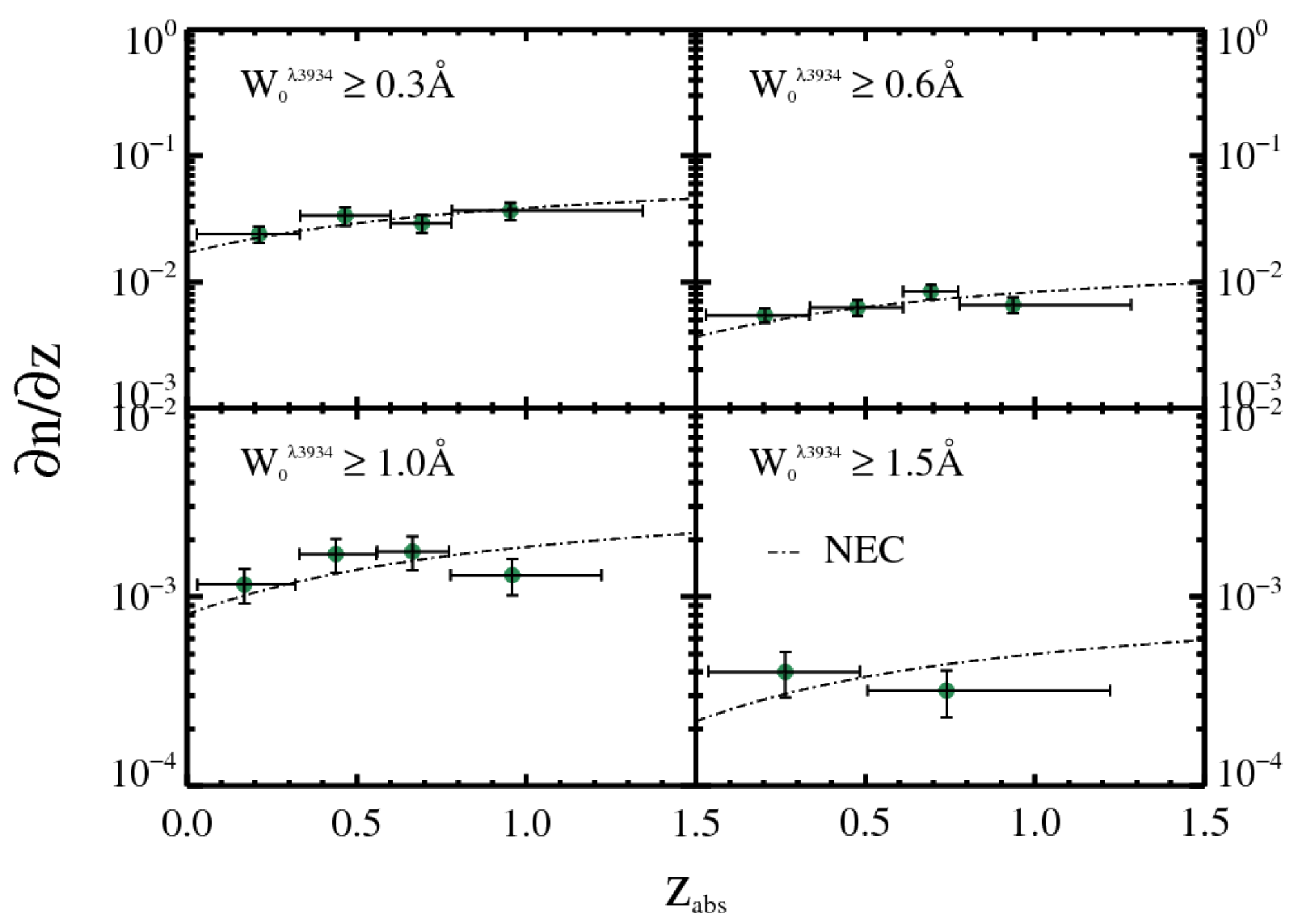} 
\caption{The \CaII\ redshift number
density as a function of $z_{abs}$ for various REW thresholds $W_0^{\lambda 3934} \ge
W_0^{min} $. The errors are determined using Poisson statistics. The bin sizes are such that there
are approximately equal numbers of systems in each bin. The no evolution curves (NEC) 
 are shown as dash-dot lines. The NECs are normalized to minimize the sum
of squared deviations of the binned data from the curve. 
With the exception of the $W_0^{\lambda3934} \ge 1.5~\mathrm{\AA}$
sample, the NECs are consistent with the data at a $ > 99.9$\% significance level. The $W_0^{\lambda3934} \ge
1.5~\mathrm{\AA}$ sample has too few data points to allow for a meaningful interpretation.}
\label{dndz} \end{figure*}

\noindent
We reiterate that
one can ignore the errors in $g(W,z)$ since the dominant contribution to the error budget comes
from the number counts, as discussed in \S2.2. The incidence of absorption lines represents the product of the
integrated number density of absorbers per co-moving volume and their effective cross section. 
Figure \ref{dndz} shows the \CaII~incidence for various equivalent width thresholds,
$\mathrm{W_0^{min} = 0.3, 0.6, 1.0, 1.5~\AA}$. In each panel, the data are binned to have 
approximately equal numbers of systems. In each panel, the dash-dot lines show
the no evolution curves (NECs) predicted by the standard cosmology (Equation \ref{dndzequation}). 
The normalization, $n_0$, has been adjusted to minimize the sum of  squared deviations of the 
binned data points from the curve. For $W_0^{\lambda3934} \ge 0.3$ \AA, the normalization 
constant, which is also the extrapolated incidence at $z=0$, is $n_0=0.017 \pm 0.001$.  
Except for the case where $W_0^{\lambda3934} \ge 1.5~\mathrm{\AA}$, 
the data are consistent with the NEC at better than the $99.9\%$ confidence level.

\begin{equation}
\frac{dn}{dz} = n_0 \frac{(1+z)^{2}}{\sqrt{\Omega_{M}(1+z)^{3} + \Omega_{\Lambda}}}
\label{dndzequation}\end{equation}

\subsection{The \CaII~ Doublet Ratio}

\begin{figure*} 
\includegraphics[width=0.75\textwidth]{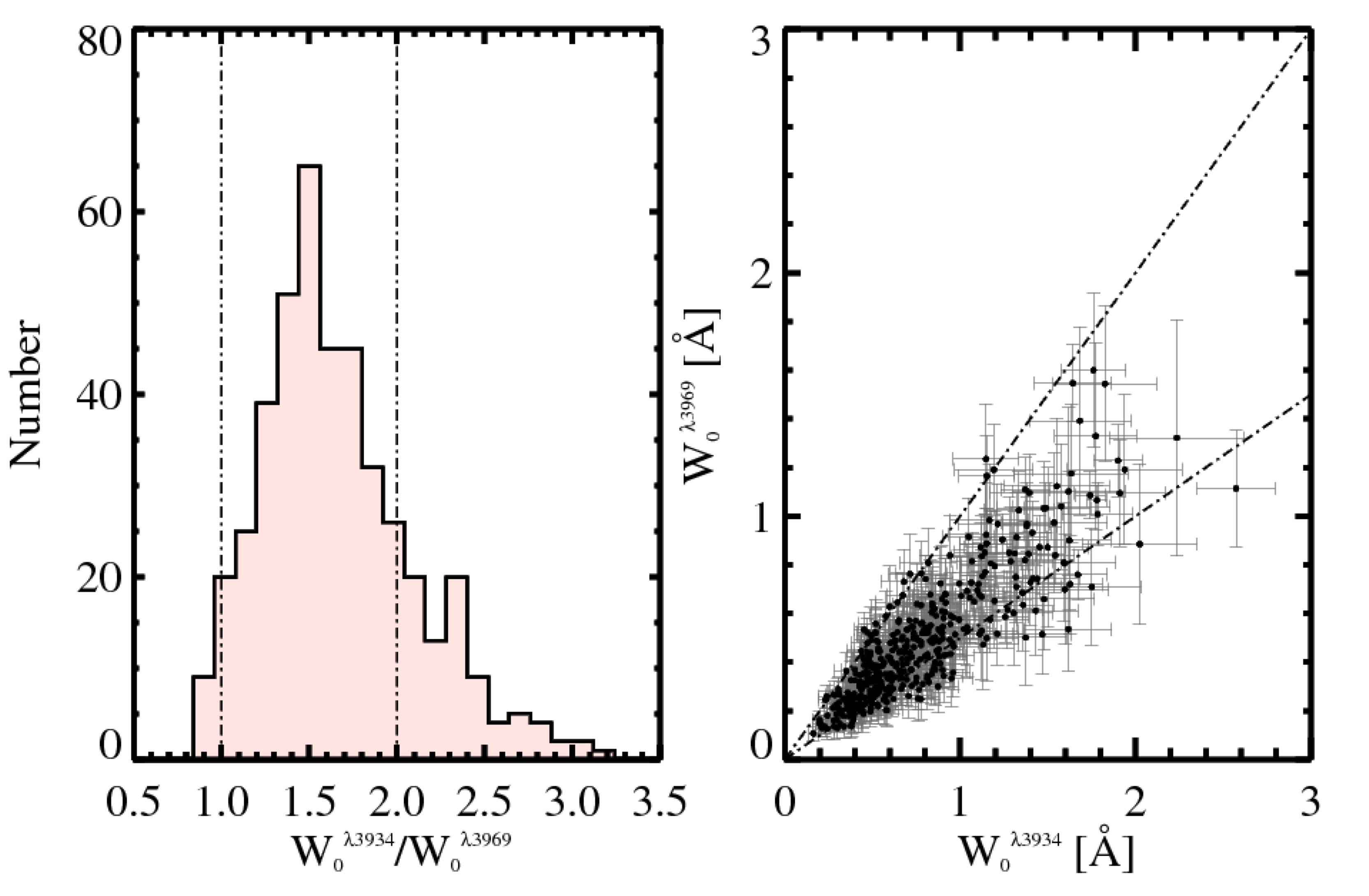} 
\caption{\textit{Left:} The
$W_0^{\lambda3934}/W_0^{\lambda3969}$~doublet ratios for the \CaII~sample. The dash-dot lines
mark the limits of $1.0$ for completely saturated systems and $2.0$ for completely unsaturated
systems. Values above and below these limits are due to poorer signal-to-noise ratio 
data. Our sample is not dominated by
either extreme DR values. \textit{Right:} $W_0^{\lambda3934}$~vs.~$W_0^{\lambda3969}$.}
\label{doubletratio} \end{figure*}

\begin{figure*} 
\includegraphics[width=0.75\textwidth]{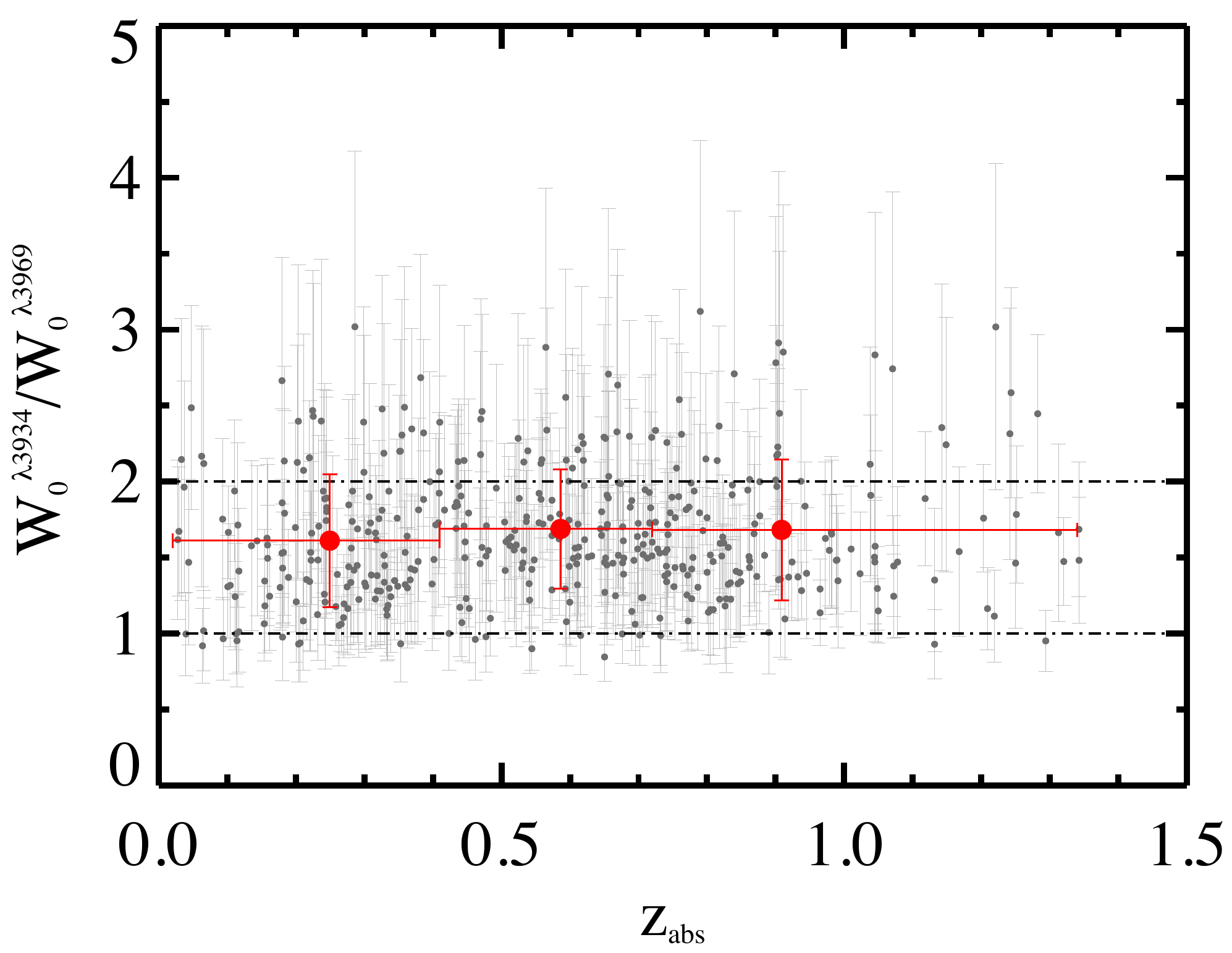} 
\caption{The
$W_0^{\lambda3934}/W_0^{\lambda3969}$ doublet ratio as a function of redshift for \CaII~systems.
There is no detected redshift evolution in the doublet ratio.} 
\label{doubletratioEvol}
\end{figure*} 

For transitions with different oscillator strengths (e.g., absorption doublets), a measured
doublet ratio (DR) is an important indicator of the degree of saturation of an absorption line.
The equivalent widths of weak unsaturated lines provide direct measurements of column densities.
For strong, saturated doublets, such as \MgII~$\lambda\lambda2796,2803$~and
\FeII~$\lambda\lambda2586,2600$, equivalent width measurements are more appropriately related to
gas velocity spreads. The observed \CaII~DRs ($W_{0}^{\lambda 3934}/W_{0}^{\lambda 3969}$) for
our sample range from $\sim 2$ for completely unsaturated systems to $\sim1$ for completely
saturated systems. The left panel in Figure \ref{doubletratio} shows the DR distribution for our
sample. It has a mean of $\sim 1.7 $ and a spread of $\sigma_{DR} \sim 0.4$. Hence, the
\CaII~doublets are on average between the two extreme possible values. The right panel in Figure
\ref{doubletratio} shows $W_0^{\lambda3969}$ versus $W_0^{\lambda3934}$ and includes the errors
on these observed values, with the dash-dot lines bounding the physically allowed DR ranges, as
in the left panel. Figure \ref{doubletratioEvol} shows the DRs as a function of redshift, along
with the propagated DR errors assuming Gaussian error distributions. There is no detectable
evolution in the DR distribution.

\subsection{The \CaII~versus \MgII~Incidence}

\begin{figure*}
\includegraphics[width=0.75\textwidth]{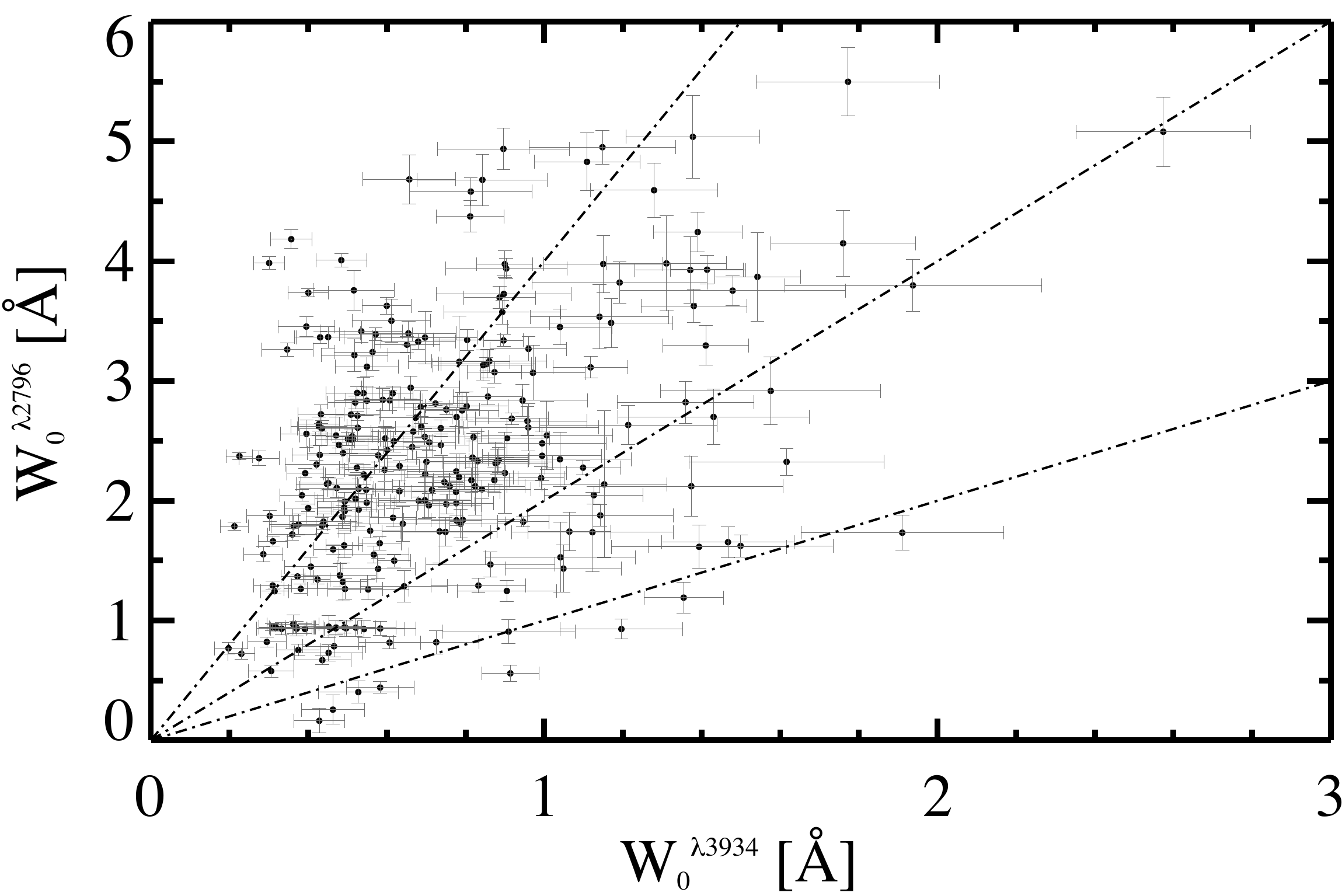} 
\caption{$W_0^{\lambda2796}$ versus
$W_{0}^{\lambda3934}$ for the 251 \CaII\ systems in our sample with detected \MgII.~ 
See text.  There is a correlation between the REWs of \MgII\ with \CaII, albeit with a large spread, but a 
sharp lower bound. The three
dash-dot lines have  $W_0^{\lambda2796}/W_{0}^{\lambda3934} = [1.0, 2.0, 4.0]$.}
\label{CaIIvsMgIIREW} \end{figure*}

\begin{figure*} 
\includegraphics[width=0.75\textwidth]{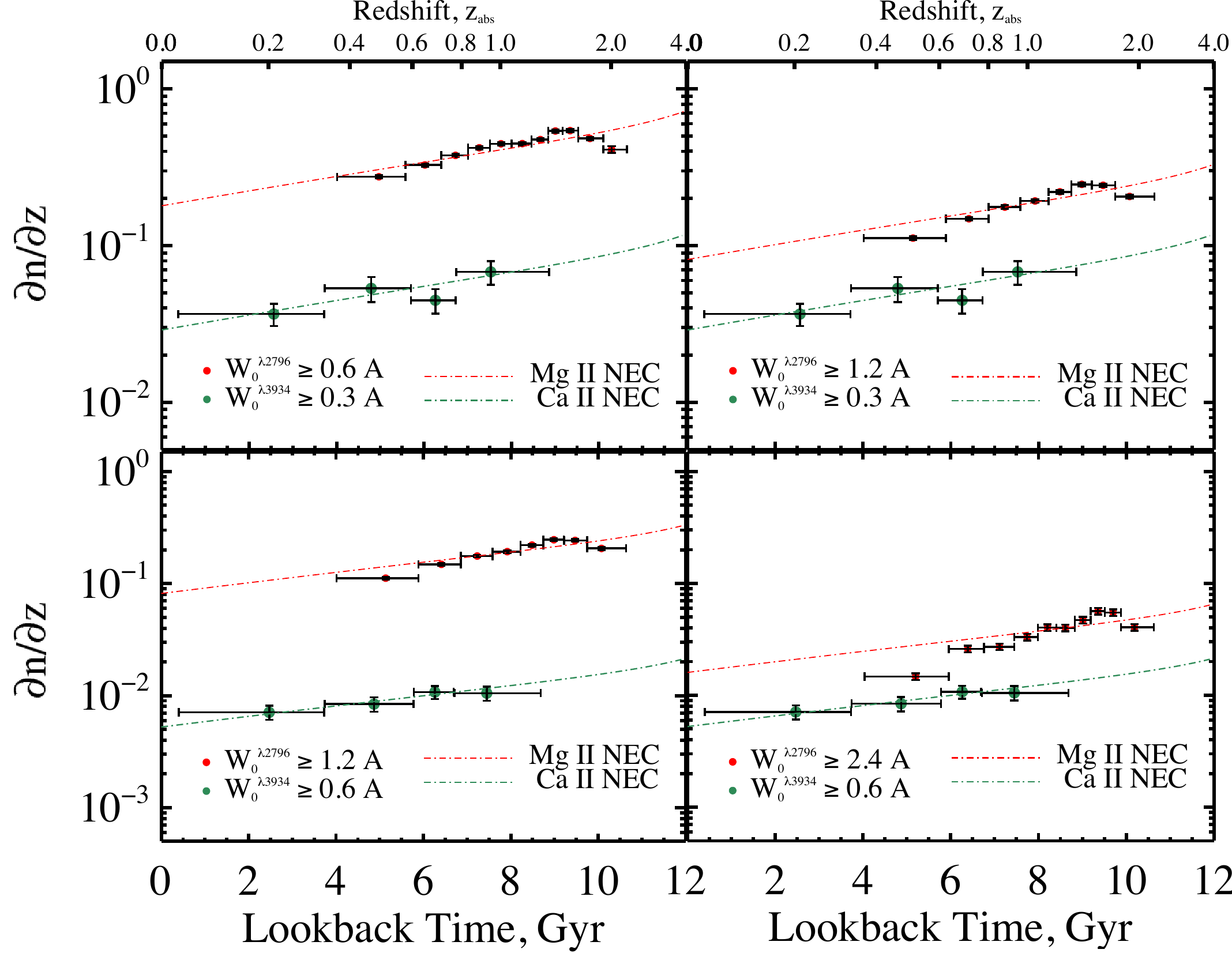} 
\caption{The
\CaII~incidence as a function of the $z_{abs}$ across various $W_0^{\lambda3934}$ thresholds.
The no evolution curve is shown as dash dot lines. The errors are propagated using the Poisson
errors. } 
\label{CaIIvsMgII} \end{figure*}

To make appropriate comparisons of the \MgII~incidence to that of \CaII, we first determine which
of the \CaII~systems in our sample have confirmed \MgII\ measurements. We made use of the data available from
the University of Pittsburgh SDSS DR4 \MgII~Catalog (Quider et al. 2011), extended up to SDSS DR7 (E. Monier, private commmunication). In
total, the extended \MgII~catalog contains over $29, 000$ doublets, which includes an additional
$\sim13, 000$ unique \MgII~systems from SDSS DR7. The Quider et al. (2011) \MgII\ sample was
selected based on $5\sigma$~and~$3\sigma$ significance cuts  for $W_0^{\lambda
2796}$ and $W_0^{\lambda 2803}$, respectively. A similar doublet ratio cut of $1.0-\sigma_{DR} \le
W_0^{\lambda2796}/W_0^{\lambda2803} \le 2.0 + \sigma_{DR}$ was also imposed to construct the
final catalog. For absorbers from SDSS DR9, we measured the
strengths of the \MgII\ doublets, and made the appropriate cuts. Note that to find \MgII, the
\CaII~system must be at $z_{abs} \gtrsim 0.4$. However, detecting the corresponding \MgII\ lines
also depends on the quasar
emission redshift and the SNR of the spectrum at the predicted \MgII\ location. Lines that
fell in the Ly$\alpha$ forest were not measured. After implementing the required selection cuts,
we have $251$ \CaII-\MgII~systems. The results are shown in Figure \ref{CaIIvsMgIIREW}. We see a
positive correlation between the strengths of the two lines, albeit with a spread that is 
quite large. However, the distribution does appear to have a sharp lower bound. The three 
dash-dot lines have $W_{0}^{\lambda2796}/W_{0}^{\lambda 3934} = [1, 2, 4]$. 

We also performed the reverse search where we looked for \CaII\ lines corresponding to \MgII\ systems from DR7
in the $0.4\le z \le1.34$ redshift interval. Only 3\% of \MgII\ systems were found to have \CaII, 
confirming that it is rare to identify \CaII~in quasar absorption-line surveys. 

We now compare the incidence of \CaII~absorbers to the more common \MgII~systems. In Figure
\ref{CaIIvsMgII}, the incidence of \CaII~is shown as the green data points
using the following REW thresholds: $W_{0}^{\lambda 3934} \ge 0.3 \mathrm{\AA}$ shown on the top
panels, and $W_{0}^{\lambda 3934} \ge 0.6 \mathrm{\AA}$ shown by the bottom panels. 
The \MgII~incidence is shown in red. The errors are derived using Poisson counting
statistics. Within each panel, we binned the data so that each point has roughly the same
number of systems. Note that these are plotted against $t_{LB}$ in the linear
scale instead of $z_{abs}$ in order to highlight the length of cosmic time that \MgII\ 
cannot trace using ground-based optical observations. Motivated by the
observed ratios in Figure \ref{CaIIvsMgIIREW}, we have chosen the following \MgII-\CaII~ratios:
$W_{0}^{\lambda 2796}/W_{0}^{\lambda 3934} = [2,4]$ for this comparison. 
For both absorbers, the NECs
were normalized to minimize the sum of the squares of the residuals. The resulting fits are
consistent with the data at $\gtrsim99\%$ confidence level for both absorbers at all REW thresholds.
The \CaII~incidence is in a sense similar to \MgII~in that the gaseous cross-sections do not
show evidence for evolution at $z_{abs} > 0.4$, and with this new result from \CaII, we extend
the same conclusions down to $z=0$. Figure \ref{CaIIvsMgII} also underscores the
rareness of \CaII~absorbers relative to \MgII. More specifically, in the left panels of Figure
\ref{CaIIvsMgII}, where $W_{0}^{\lambda 2796}= 2W_{0}^{\lambda 3934}$, the incidence of \CaII~is
roughly a factor of $ \sim 10$ times smaller. At the larger REW ratios (right panels), as the
number of stronger \MgII~absorber becomes rare, this fraction drops to a factor of $ \sim 3$~to~$4$.

\section{Investigating the Possibility of Two \CaII~Absorber Populations}

The $W_0^{\lambda3934}$~distribution shown in Figure \ref{truerewdist} reveals a break in 
$\partial n/\partial W_0^{\lambda3934}$ at $W_0^{\lambda3934}=0.88~\mathrm{\AA}$. The need for 
a strong and a weak component to adequately fit the overall distribution (Equation 4) suggests that we 
should investigate trends which might further reveal the properties of these components. Below we 
search for identifiable trends based on: (1) the $W_0^{\lambda3934}$ value and \CaII~DR, and 
(2) the $W_0^{\lambda3934}$ value and the \MgII-to-\CaII~ratio ($W_0^{\lambda2796}/W_0^{\lambda3934}$).

\subsection{Trends with $W_0^{\lambda3934}$ and \CaII~DR}

Here we explore the possible role of the \CaII~DR in isolating the two components of the 
$W_0^{\lambda3934}$ distribution. To do this we divide the entire \CaII~sample into four subsamples 
of roughly equal size
based on their DRs and $W_0^{\lambda3934}$ values. This can be accomplished by making divisions above and 
below DR$~=1.5$ and $W_0^{\lambda3934}=0.7$ \AA. Note that these values lie close the the mean doublet 
ratio of the entire sample and the location of the break
in Figure \ref{truerewdist}. 

\begin{figure*}
\includegraphics[width=0.75\textwidth]{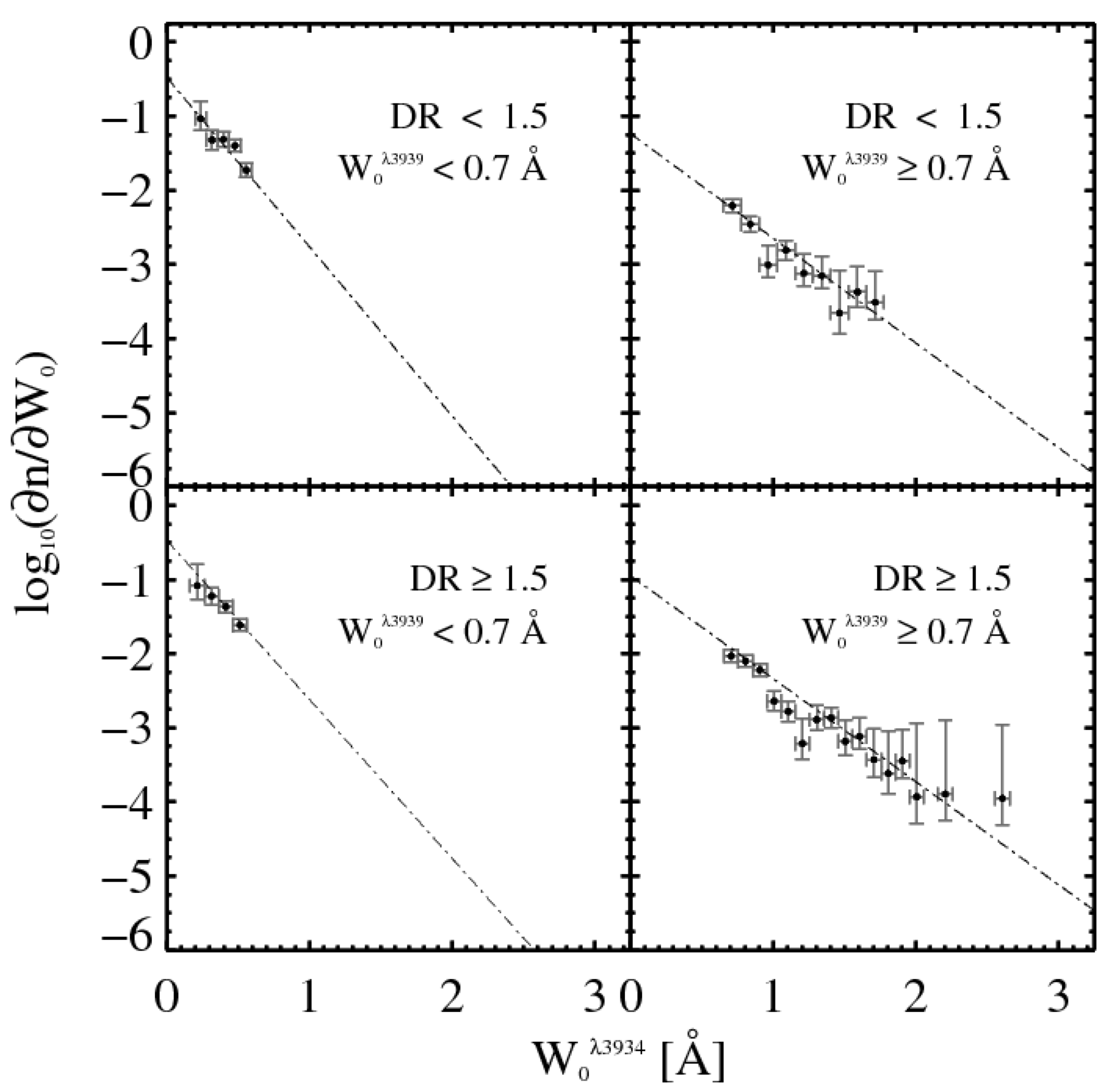}
\caption{The sensitivity-corrected equivalent with distributions for four roughly equal subsamples of 
\CaII~absorbers divided according to \CaII~DR and $W_0^{\lambda3934}$. The dash-dot lines are the MLE 
single power-law fits.}
\label{FourGroups}
\end{figure*}

In Figure \ref{FourGroups}, we plot the sensitivity-corrected 
$W_0^{\lambda3934}$ distributions for the four defined subsamples. The resulting four distributions can now be 
accurately parametrized by single exponential functions. The best-fit single exponential functions to the 
unbinned data are shown as dash-dot lines. All four single exponential functions fit the data in their corresponding
subsamples at a better than $99\%$ confidence level.
Moreover, the MLE slopes for both subsamples with $W_0^{\lambda3934} < 0.7$ \AA\ (left panels) are 
consistent with the $W_{wk}^{\star}$ value for the overall sample to within the errors, 
and the MLE slopes for both subsamples with $W_0^{\lambda3934} > 0.7$ \AA\ (right panels) are 
consistent with the $W_{str}^\star$ value to within the errors. 
Also, dividing the entire sample at DR=1.5 into
two subsamples yielded the same two component distribution specified in Equation 4 to within the errors. 

Thus, at that accuracy of our data, the \CaII~DR alone does not play a role in separating the \CaII~absorbers 
into two populations. A series of KS tests were also performed and found to support this conclusion.

\subsection{Trends with $W_0^{\lambda3934}$ and $W_0^{\lambda2796}/W_0^{\lambda3934}$}

For almost all \CaII~absorbers with redshifts $z>0.4$ we have information on the corresponding 
\MgII~absorption. Therefore, we can explore if \MgII~information can be used to isolate the two components 
of the $W_0^{\lambda3934}$ distribution. From past work we know that \MgII~absorption doublets found
in SDSS surveys are generally saturated (Quider et al. 2011). As explained in \S3.3, this means that \MgII~rest 
equivalent widths are more 
representative of low-ionization gas velocity spreads rather than Mg$^+$ column densities. 
However, the \CaII~doublet is generally unsaturated or only partially saturated, so to some 
degree the \CaII~rest equivalent widths must be representative of Ca$^+$ column densities. 

\begin{figure*}
\includegraphics[width=0.75\textwidth]{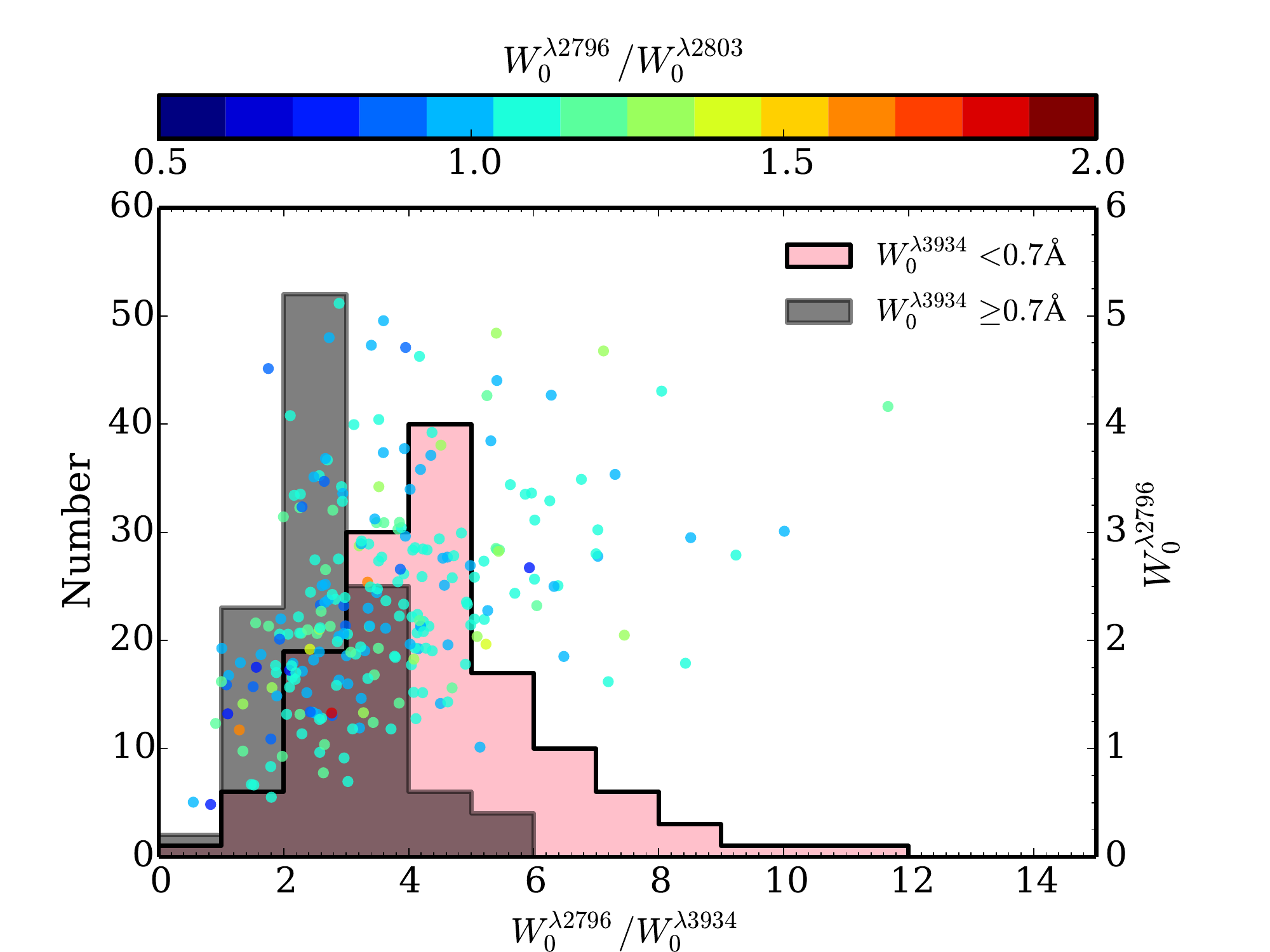}
\caption{The bimodal distribution of the $W_0^{\lambda2796}/W_0^{\lambda2803}$~ratio 
for the weak and strong \CaII~absorbers divided at $W_0^{\lambda3934} = 0.7$ \AA. 
Note that on this same figure we also plot 
$W_0^{\lambda2796}/W_0^{\lambda3934}$ (x-axis) as a function of $W_0^{\lambda2796}$ (right y-axis); these
data points are color-coded (online version only) according to the top color bar to show the saturation level of the \MgII~doublet.}
\label{mgIIcaIIratios}
\end{figure*}

In Figure \ref{mgIIcaIIratios} we plot the two observed  
$W_0^{\lambda2796}/W_0^{\lambda3934}$ ratio histograms for \CaII~absorbers with
$W_0^{\lambda3934} < 0.7$ \AA\ and $W_0^{\lambda3934} \ge 0 .7$ \AA. This separation value
is the same as the one used in \S4.1 and is again motivated by our desire to 
roughly equalize the number of systems in each of two subsamples.
There are $\sim120$ absorbers in each subsample. 

By using this $W_0^{\lambda3934}$ separation value and including \MgII~information, we produced 
Figure \ref{mgIIcaIIratios}, which shows the \CaII~absorbers to 
be a bimodal population, with the weaker \CaII~absorbers having a larger (on average) and wider spread in 
$W_0^{\lambda2796}/W_0^{\lambda3934}$ than the stronger \CaII~absorbers. 
A simple KS-test renders the two distributions inconsistent 
with one another at a $>99\%$ confidence level. 
This bimodality provides supporting evidence that stronger and 
weaker \CaII~absorbers (i.e., absorbers with relatively higher and lower Ca$^+$ column density values) may 
be different populations. We note that with the exception of a few data points, the \MgII~absorbers 
associated with the \CaII~absorbers have saturated doublets, which means that $W_0^{\lambda2796} $ values
are indicative of gas velocity spreads. See Figure \ref{mgIIcaIIratios} and its caption for 
color-coded data points (online version only) on \MgII~doublet ratios and some additional explanation.
 
\begin{figure*}
\includegraphics[width=0.75\textwidth]{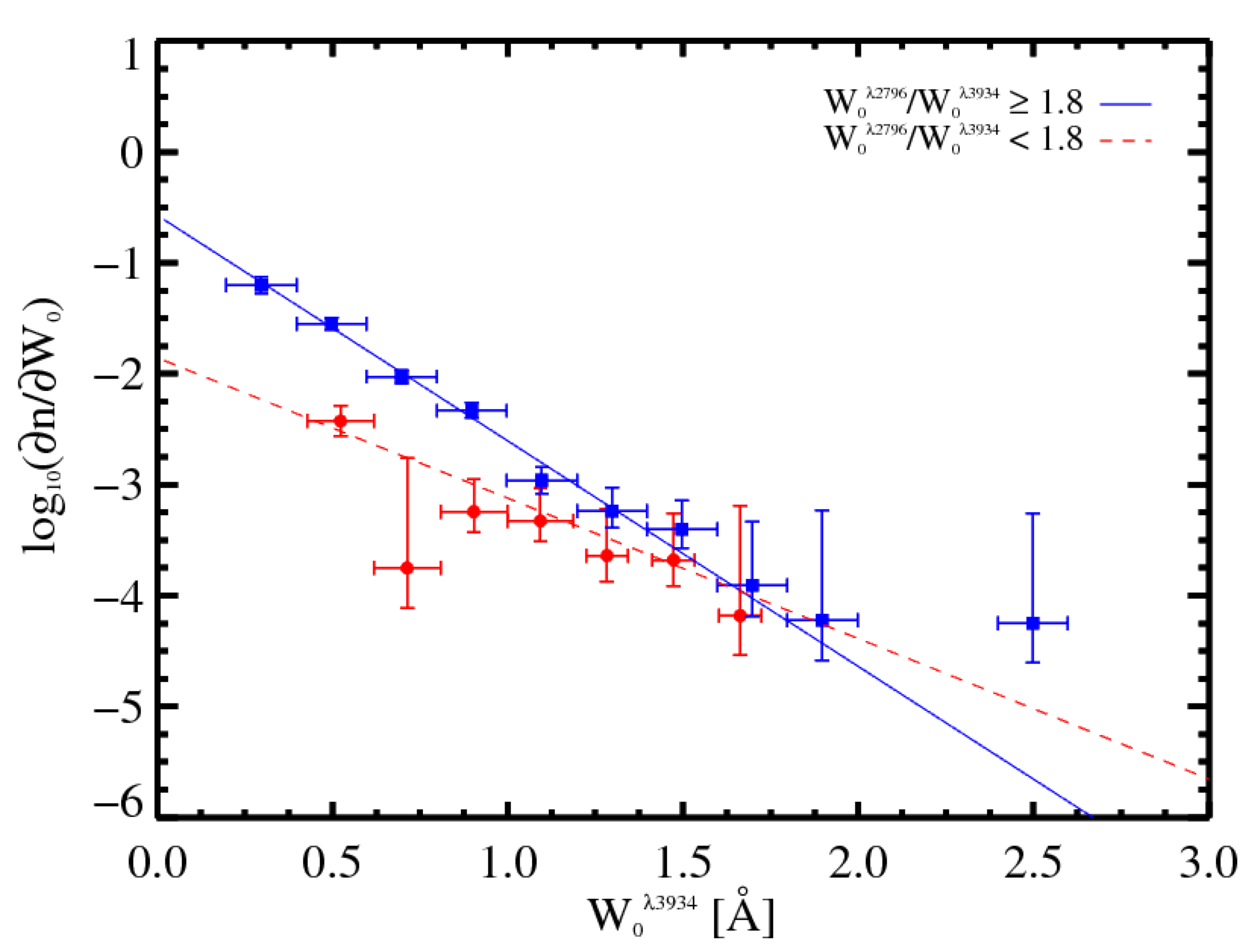}
\caption{The sensitivity-corrected equivalent width distributions for two subsamples of \CaII~absorbers separated at 
$W_0^{\lambda2796}/W_0^{\lambda3934}=1.8$. The separation into two-single power law fits 
is clear. The blue squares and red circles represent the subsamples with a  
$W_0^{\lambda2796}/W_0^{\lambda3934}$ ratio greater than and less than $1.8$, respectively. 
The solid blue and dashed red lines are the best-fit single-exponential MLE fits to the unbinned distributions.}
\label{MgIICaIIREW}
\end{figure*}

Finally, in Figure \ref{MgIICaIIREW} we show that for those 
\CaII~absorbers with \MgII~information, it is possible to separate the $W_0^{\lambda3934}$ distribution shown 
in Figure \ref{truerewdist} into two single power-law distributions over the entire range of $W_0^{\lambda3934}$ values. 
This is done by forming two subsamples divided at a $W_0^{\lambda2796}/W_0^{\lambda2803}$~ratio of $\sim 1.8$, 
but in this case the subsamples are not approximately of equal size.

 The slope of the steeper distribution is found to be consistent with the slopes of the 
 weak component of the distribution in Figure \ref{truerewdist} and the top-left and 
 bottom-left panels of Figure \ref{FourGroups} (i.e with $W_0^{\lambda3934} < 0.7~\mathrm{\AA}$). 
 Similarly, the flatter red distribution is also consistent with the corresponding results 
 for the strong systems in Figures \ref{truerewdist} and \ref{FourGroups}. 

\section{Summary and Conclusions}

We have presented the results of a blind survey for intervening \CaII~absorption-line systems 
using $\sim 95,000$ quasar spectra from the seventh and ninth data release of the SDSS. Our results represent
the largest compilation of \CaII~absorbers to date. The rest wavelengths of the 
\CaII$\lambda\lambda 3934,3969$~doublet resonance transition allow us to probe redshifts $z \lesssim 1.34$, 
which corresponds to the most recent $\sim 8.9$ Gyrs of the history of the Universe. \CaII~absorbers are 
considerably more rare than \MgII~ absorbers.
However, it is notable that with the orginal SDSS spectrograph, \MgII~absorbers at $z \lesssim 0.4$ are not accessible. 
Therefore, studies of \CaII~absorbers in quasar spectra are the only absorption-line systems which are 
generally accessible with SDSS spectra at  $z\lesssim0.4$, which is equivalent to the past $\sim 4.3$ Gyrs 
of cosmic time. Consequently, within the SDSS spectral window \CaII~presents a unique opportunity
for ground-based studies of cool, metal-rich gas around galaxies at the lowest redshifts, and such
studies can help to constrain models for the existence of cool
gas in the extended gaseous halos of galaxies.

Our blind survey resulted in the identification of $435$ \CaII~absorbers at rest equivalent width significance 
levels $\geq5\sigma$ for  $W_0^{\lambda3934}$ and $\geq2.5\sigma$ for $W_0^{\lambda3969}$, 
within the physically-allowable doublet ratio range, i.e.,  
$ 1 - \sigma_{DR} \leq W_0^{\lambda3934}/W_0^{\lambda3969} \leq 2 + \sigma_{DR}$. Of these detections,
251 \CaII~absorbers at $z\gtrsim 0.4$ were found to have associated \MgII~absorption, which is essentially all of them. 

The sensitivity-corrected $W_0^{\lambda3934}$ distribution cannot be fitted by a single-component 
exponential function, but a two-component exponential function describes the data well. 
We find $ {\partial n}/{\partial W_{0}^{\lambda 3934}}= ({N_{wk}^{\star}}/{W_{wk}^{\star}})
exp({ -{W_{0}^{\lambda 3934}}/{W_{wk}^{\star} } }) + ({N_{str}^{\star}}/{W_{str}^{\star}})
exp({ -{W_{0}^{\lambda 3934}}/{W_{str}^{\star} } })$, with
$N_{wk}^{\star}=0.140\pm0.029$, $W_{wk}^{\star}=0.165\pm 0.020~\textrm{\AA}$, 
$N_{str}^{\star}=0.024\pm0.020$, and $W_{str}^{\star}=0.427\pm 0.101~\textrm{\AA}$. 
This suggests that the \CaII~absorbers are composed of at least two distinct populations (Figure 7).

The \CaII~absorber incidence
was found to not evolve in the standard cosmology, implying that the 
product of integrated \CaII~absorber cross section and their comoving number density has remained 
roughly constant over the last $\sim 8.9$ Gyrs.
 The normalization of the no-evolution curve, which is also
the incidence extrapolated to $z=0$, is $n_0=0.017 \pm 0.001$ for the sample with 
$W_{0}^{\lambda 3934} \ge 0.3$ \AA.

Furthermore, we have demonstrated that the incidence of \CaII~absorbers relative to the 
more common \MgII~absorbers in quasar spectra is about $3$ to $10$ times smaller, depending on the REW threshold
used for the comparison (Figure 15).
 
Finally, we performed some investigations to determine if we could use available \CaII~absorber 
properties, specifically doublet ratio and \MgII\ information, to isolate the ``weak'' 
and ``strong'' populations of \CaII~absorbers. While it was not possible to do this using 
the \CaII\ doublet ratio, we did find that \MgII\ information could be used to isolate the 
two populations (Figures 16 and 17).


\nocite{*} 
\bibliography{References}{} 
\bibliographystyle{mn2e}
\section*{Acknowledgments}

GMS acknowledges support from a Zaccheus Daniel Fellowship and a Dietrich School of Arts and
Sciences Graduate Fellowship and PITT PACC Fellowship from the University of Pittsburgh.
We thank Eric Monier for providing his list of DR7 MgII absorbers to us (in order to supplement 
the DR4 list of Quider et al. (2011)) prior to publication. We thank Dan Nestor for making his continuum 
fitting software available for this work. 

Funding for SDSS has been provided by the 
Alfred P. Sloan Foundation, the Participating Institutions, the National Science Foundation, 
and the US Department of Energy Office of Science. The SDSS is managed by the Astrophysical 
Research Consortium for the Participating Institutions. The Participating Institutions are 
the American Museum of Natural History, Astrophysical Institute Potsdam, 
University of Basel, University of Cambridge, Case Western Reserve University, University of Chicago, 
Drexel University, Fermilab, the Institute for Advanced Study, the Japan Participation Group, Johns Hopkins 
University, the Joint Institute for Nuclear Astrophysics, the Kavli Institute for Particle Astrophysics 
and Cosmology, the Korean Scientist Group, the Chinese Academy of Sciences (LAMOST), Los 
Alamos National Laboratory, the Max-Planck-Institute for Astronomy (MPIA), the 
Max-Planck-Institute for Astrophysics (MPA), New Mexico State University, Ohio State University, 
University of Pittsburgh, University of Portsmouth, Princeton University, 
the United States Naval Observatory, and the University of Washington. 

\end{document}